\documentclass{aa}
\usepackage{natbib}
\bibpunct{(}{)}{;}{a}{}{,}
\usepackage{graphicx} 

\newlength{\zero}
\newcommand{\wz}{\hspace{\zero}}
\newcommand{\mb}[1]{\mbox{\boldmath $#1$}}
\newcommand{\mbsm}[1]{\mbox{\scriptsize \boldmath $#1$}}
\newcommand{\ip}{i+{1\over2}}
\newcommand{\im}{i-{1\over2}}
\newcommand{\jp}{j+{1\over2}}
\newcommand{\jm}{j-{1\over2}}

\newenvironment{equationarray}
{\arraycolsep 0.14 em
\begin {eqnarray}}
{\end {eqnarray}}

\newenvironment{equationarray*}
{\arraycolsep 0.14 em
\begin {eqnarray*}}
{\end {eqnarray*}}

\begin{document}

\title{Relativistic simulations of rotational core collapse. \\
I. Methods, initial models, and code tests}

\author{Harald Dimmelmeier \inst{1} 
\and Jos\'e A. Font \inst{1,2} 
\and Ewald M\"uller \inst{1}}

\offprints{Harald Dimmelmeier, \\ \email{harrydee@mpa-garching.mpg.de}}

\institute{Max-Planck-Institut f\"ur Astrophysik,
Karl-Schwarzschild-Str. 1, 85741 Garching, Germany
\and
Departamento de Astronom\'{\i}a y Astrof\'{\i}sica,
Universidad de Valencia, 46100 Burjassot (Valencia), Spain}

\date{Received date / Accepted date}

%%%%%%%%%%%%%%%%%%%%%%%%%%%%%%%%%%%%%%%%%%%%%%%%%%%%%%%%%%%%%%%%%%%%%%%%%%%%%%%%%%%
%%%%%%%%%%%%%%%%%%%%%%%%%%%%%%%%%%%%%%%%%%%%%%%%%%%%%%%%%%%%%%%%%%%%%%%%%%%%%%%%%%%
%%%%%%%%%%%%%%%%%%%%%%%%%%%%%%%%%%%%%%%%%%%%%%%%%%%%%%%%%%%%%%%%%%%%%%%%%%%%%%%%%%%

\abstract{

  We describe an axisymmetric general relativistic code for rotational
  core collapse. The code evolves the coupled system of metric and
  fluid equations using the ADM 3\,+\,1 formalism and a conformally
  flat metric approximation of the Einstein equations. Within this
  approximation the ADM 3\,+\,1 equations reduce to a set of five
  coupled non-linear elliptic equations for the metric components.
  The equations are discretized on a 2D grid in spherical polar
  coordinates and are solved by means of a Newton-Raphson iteration
  using a block elimination scheme to solve the diagonally dominant,
  sparse linear system arising within each iteration step. The
  relativistic hydrodynamics equations are formulated as a first-order
  flux-conservative hyperbolic system and are integrated using
  high-resolution shock-capturing schemes based on Riemann solvers. We
  assess the quality of the conformally flat metric approximation for
  relativistic core collapse and present a comprehensive set of tests
  which the code successfully passed. The tests include relativistic
  shock tubes, the preservation of the rotation profile and of the
  equilibrium of rapidly and differentially rotating neutron stars
  (approximated as rotating polytropes), spherical relativistic core
  collapse, and the conservation of rest-mass and angular momentum in
  dynamic spacetimes. The application of the code to relativistic
  rotational core collapse, with emphasis on the gravitational waveform
  signature, is presented in an accompanying paper.

  \keywords{Gravitation -- Gravitational waves -- Hydrodynamics --
    Methods: numerical -- Relativity} 
}

\authorrunning{H.~Dimmelmeier et al.}
\titlerunning{Relativistic rotational core collapse. I}

\maketitle

%%%%%%%%%%%%%%%%%%%%%%%%%%%%%%%%%%%%%%%%%%%%%%%%%%%%%%%%%%%%%%%%%%%%%%%%%%%%%%%%%%%
%%%%%%%%%%%%%%%%%%%%%%%%%%%%%%%%%%%%%%%%%%%%%%%%%%%%%%%%%%%%%%%%%%%%%%%%%%%%%%%%%%%
%%%%%%%%%%%%%%%%%%%%%%%%%%%%%%%%%%%%%%%%%%%%%%%%%%%%%%%%%%%%%%%%%%%%%%%%%%%%%%%%%%%

\section{Introduction}
\label{sec:introduction}

The advent of gravitational wave astronomy -- major interferometric
gravitational wave detectors have already started taking data --
aggravates the need for gravitational wave templates computed from
theoretical models of potential sources of gravitational radiation
\citep{pradier_01_a}. Among the most promising sources are aspherical
core collapse supernovae. However, reliable core collapse simulations,
in particular those of rotational core collapse, require a
relativistic treatment of the dynamics because of the counteracting
stabilizing effect of rotation and the destabilizing effect of
relativistic gravity. In addition, the simulations should incorporate
a realistic equation of state (EoS) and an accurate treatment of
neutrino transport. 

Due to these complexities, all previous investigations have either
neglected or approximated one or all of these requirements (see,
e.g.\ \citet{mueller_98_a} and references therein). Up to now there
exist no successful multidimensional numerical simulations of
rotational core collapse to neutron stars in general relativity, even
for simple matter models. On the other hand, simulations which include
better microphysics in the form of realistic (nuclear) equations of
state or neutrino transport have either been confined to
spherical symmetry or have used Newtonian gravity. 

There is an obvious need to improve this situation if one wants to
answer questions such as: What is the quantitative influence of
general relativity on the dynamics of core collapse compared to
Newtonian gravity? How do higher densities and velocities expected in
relativistic simulations change the properties of the rotating
proto-neutron star? To what extent do relativistic effects alter the
gravitational wave signal? How does the rotation rate evolve in a
relativistic simulation, and what consequences does that have on the
development of dynamical or secular instabilities in the neutron star?

To this end we have started a project to compute reliable
gravitational wave templates from rotational core collapse and
nonspherical core collapse supernovae. In a first step we have
improved the treatment of the dynamics beyond the still commonly used
Newtonian approach and computed the general relativistic hydrodynamics
of rotational core collapse consistently with the evolution of the
axisymmetric dynamic spacetime. The computational tool developed by us
for this purpose is described in this publication, while its
application to the gravitational collapse of rotating stellar cores
approximated as polytropes is devoted to an accompanying paper
\citep{dimmelmeier_02_b}. We note that in \citet{dimmelmeier_01_a} we
have already presented preliminary results of our investigation.
General relativistic simulations of improved models including both a
realistic equation of state and an approximate treatment of weak
interactions and neutrino transport are planned in the near future.

The asumptions we adopt in the present and accompanying paper are as
follows: Our matter model obeys an ideal gas EoS with the pressure
$ P $ consisting of a polytropic and a thermal part \citep{janka_93_a,
zwerger_97_a}. Since we are mostly interested in the gravitational
radiation emission, which is controlled by the bulk motion of the
fluid, we neglect any other microphysics like electron capture and
neutrino transport. The initial configurations are rotating
relativistic polytropes in equilibrium \citep{komatsu_89_a,
komatsu_89_b, stergioulas_95_a} which are marginally stable. Their
collapse is initiated by decreasing the effective adiabatic index
$ \gamma_{\rm p} $ to some prescribed fixed value. Our initial data
solver allows us to construct both uniformly and differentially
rotating polytropes.

We solve the general relativistic hydrodynamic equations
numerically. The metric equations are approximated by adopting the
so-called {\em conformal flatness condition} for the three-metric
$ \gamma_{ij} $ \citep{wilson_96_a}. This results in a constrained
system of elliptic metric equations which are much simpler than the
full system of Einstein equations. The dimensionality of the problem
is reduced by assuming both axisymmetry with respect to the rotation
axis, and equatorial plane symmetry. The numerical code uses a
spherical grid in spherical polar $ (r, \theta) $ coordinates.

This paper is organized as follows: In
Section~\ref{sec:mathematical_framework} we present the mathematical
framework used in our approach, namely the conformally flat metric
equations and the relativistic hydrodynamic equations.
Section~\ref{sec:initial_models} is devoted to describing the initial
models. Next, in Section~\ref{sec:numerical_implementation} we
describe all relevant issues concerning our numerical
implementation. Section~\ref{sec:tests} presents a comprehensive
number of tests we have used to validate our code. The paper ends with
a summary in Section~\ref{sec:conclusions}. 

Unless otherwise stated we use geometrized units ($ G = c = 1 $).
Greek (Latin) indices run from 0 to 3 (1 to 3). Latin indices are
raised and lowered using the three-metric.

%%%%%%%%%%%%%%%%%%%%%%%%%%%%%%%%%%%%%%%%%%%%%%%%%%%%%%%%%%%%%%%%%%%%%%%%%%%%%%%%%%%
%%%%%%%%%%%%%%%%%%%%%%%%%%%%%%%%%%%%%%%%%%%%%%%%%%%%%%%%%%%%%%%%%%%%%%%%%%%%%%%%%%%
%%%%%%%%%%%%%%%%%%%%%%%%%%%%%%%%%%%%%%%%%%%%%%%%%%%%%%%%%%%%%%%%%%%%%%%%%%%%%%%%%%%

\section{Mathematical framework}
\label{sec:mathematical_framework}

%%%%%%%%%%%%%%%%%%%%%%%%%%%%%%%%%%%%%%%%%%%%%%%%%%%%%%%%%%%%%%%%%%%%%%%%%%%%%%%%%%%
%%%%%%%%%%%%%%%%%%%%%%%%%%%%%%%%%%%%%%%%%%%%%%%%%%%%%%%%%%%%%%%%%%%%%%%%%%%%%%%%%%%

\subsection{Metric equations}
\label{subsec:metric_equations}

We adopt the ADM 3\,+\,1 formalism \citep{arnowitt_62_a} to
foliate the spacetime into a set of non-intersecting spacelike
hypersurfaces. The line element reads
\begin{equation}
  ds^2 = - \alpha^2 dt^2 + \gamma_{ij} (dx^i + \beta^i dt)
  (dx^j + \beta^j dt),
  \label{eq:line_element}
\end{equation}
with $ \alpha $ being the lapse function, which describes the rate
of advance of time along a timelike unit vector $ n^{\mu} $ normal
to a hypersurface, $ \beta^i $ being the spacelike shift three-vector
which describes the motion of coordinates within a surface, and
$ \gamma_{ij} $ being the spatial three-metric.

In the 3\,+\,1 formalism, the Einstein equations split into evolution equations
for the three-metric $ \gamma_{ij} $ and the extrinsic curvature
$ K_{ij} $, and constraint equations which must be fulfilled at every
spacelike hypersurface:
\begin{equationarray}
  \partial_t \gamma_{ij} & = & - 2 \alpha K_{ij} + \nabla_i \beta_j +
  \nabla_j \beta_i,
  \label{eq:adm_metric_equation_1} \\
  \partial_t K_{ij} & = &  - \nabla_i \nabla_j \alpha + 
  \alpha (R_{ij} + K K_{ij} - 2 K_{ik} K_j^k) 
  \nonumber \\ 
  & & + \beta^k \nabla_k K_{ij} + K_{ik} \nabla_j \beta^k + K_{jk}
  \nabla_i \beta^k
  \nonumber \\ && -
  8 \pi \alpha \! \left( \! S_{ij} - \frac{\gamma_{ij}}{2}
  (S - \rho_{\rm H}) \right)\!,
  \label{eq:adm_metric_equation_2} \\
  0 & = & R + K^2 - K_{ij} K^{ij} - 16 \pi \rho_{\rm H},
  \label{eq:adm_metric_equation_3} \\
  0 & = & \nabla_i (K^{ij} - \gamma^{ij} K) - 8 \pi S^j.
  \label{eq:adm_metric_equation_4}
\end{equationarray}%
In these equations $ \nabla_i $ is the covariant derivative with
respect to the three-metric $ \gamma_{ij} $, $ K_{ij} $ is the
extrinsic curvature, $ R_{ij} $ is the Ricci tensor, $ R $ is the
scalar curvature and $ K = K^i_i $ is the trace of the extrinsic
curvature. The matter fields appearing in the above equations,
$ S_{ij} $, $ S^j $ and $ \rho_{\rm H} $, are the spatial components
of the stress-energy tensor, the momenta and the total energy,
respectively.

Following \citet{wilson_96_a} (see also \citet{flanagan_99_a,
mathews_00_a}) we approximate the general metric $ g_{\mu \nu} $ by
replacing its spatial three-metric $ \gamma_{ij} $ with the
conformally flat (CF) three-metric (conformal flatness condition --
CFC hereafter):
\begin{equation}
  \gamma_{ij} = \phi^4 \hat{\gamma}_{ij},
  \label{eq:cfc_three_metric}
\end{equation}
where $ \hat{\gamma}_{ij} $ is the flat metric
($ \hat{\gamma}_{ij} = \delta_{ij} $ in Cartesian coordinates). In
general, the conformal factor $ \phi $ depends on the coordinates
$ (t,r,\theta) $. Therefore, at all times during a numerical
simulation we assume that all off-diagonal components of the
three-metric are zero, and the diagonal elements have the common
factor $ \phi^4 $. Note that all metric quantities with a hat are
defined with respect to the flat three-metric $ \hat{\gamma}_{ij} $.

Within this approximation the ADM equations reduce to a set of five
coupled elliptic (Poisson-like) equations for the metric components,
\begin{equationarray}
  \hat{\Delta} \phi & = & - 2 \pi \phi^5
  \left( \rho h W^2 - P + \frac{K_{ij} K^{ij}}{16 \pi} \right),
  \label{eq:metric_equation_1} \\
  \hat{\Delta} (\alpha \phi) & = & 2 \pi \alpha \phi^5
  \left( \rho h (3 W^2 - 2) + 5 P + \frac{7 K_{ij} K^{ij}}{16 \pi} \right),
  \label{eq:metric_equation_2} \\
  \hat{\Delta} \beta^i & = & 16 \pi \alpha \phi^4 S^i + 2 \hat{K}^{ij}
  \hat{\nabla}_j \left( \frac{\alpha}{\phi^6} \right) - \frac{1}{3}
  \hat{\nabla}^i \hat{\nabla}_k \beta^k,
  \label{eq:metric_equation_3}
\end{equationarray}%
where $ \hat{\nabla} $ and $ \hat{\Delta} $ are the flat space Nabla
and Laplace operator, respectively. The transformation behavior
between the extrinsic curvature defined on $ \gamma_{ij} $ and
$ \hat{\gamma}_{ij} $ is as follows:
\begin{equation}
  K_{ij} = \phi^{-2} \hat{K}_{ij}, \qquad K^{ij} = \phi^{-10} \hat{K}^{ij}.
  \label{eq:extrinsic_curvature_transformation}
\end{equation}

The metric
equations~(\ref{eq:metric_equation_1}--\ref{eq:metric_equation_3})
couple to each other via their right hand sides, and
in case of the equations for $ \beta^i $ via the operator
$ \hat{\Delta} $ acting on the vector $ \beta^i $. The equations are
dominated by the source terms involving the hydrodynamic quantities
$ \rho $, $ P $, and $ v^i $, whereas the nonlinear coupling through
the other, purely metric, source terms becomes only important for
strong gravity. On each time slice the metric is solely
determined by the instantaneous hydrodynamic state, i.e.\ the
distribution of matter in space. 

Applying the CFC to the traceless part of
Eq.~(\ref{eq:adm_metric_equation_1}) yields the following
relation between the conformal factor and the shift vector components:
\begin{equation}
  \partial_t \phi = \frac{\phi}{6} \nabla_k \beta^k.
  \label{eq:phi_t_relation}
\end{equation}
With this, Eq.~(\ref{eq:adm_metric_equation_1}) transforms into a
definition for the extrinsic curvature components:
\begin{equation}
  K_{ij} = \frac {1}{2 \alpha} \left( \nabla_i \beta_j +
    \nabla_j \beta_i - \frac{2}{3} \gamma_{ij} \nabla_k \beta^k \right).
  \label{eq:definition_of_extrinsic_curvature}
\end{equation}

The above approximation is exact in spherical symmetry. Nevertheless
we have used it for our simulations of rotational core collapse. The
rationale behind this choice is the well-justified assumption that the
energy emitted in form of gravitational waves is only about
$ 10^{-6} $ of the energy released during gravitational collapse in a
Type~II supernova event \citep{mueller_82_a, moenchmeyer_91_a,
zwerger_97_a}. Furthermore, during core collapse deviations from
spherical symmetry are modest even for rapidly rotating cores compared
to those occuring in a neutron star merger \citep{wilson_96_a}). This
fact is also reflected in the strength of the gravitational wave
signal from core collapse, which is three orders of magnitude smaller
than that resulting from the plunge phase of a compact binary. In
Section~\ref{subsec:cfc_quality} we will discuss the quality of the CFC
in the context of rotational core collapse in more detail.

As a consequence of setting the off-diagonal elements of
$ \gamma_{ij} $ to zero, the degrees of freedom representing
gravitational waves are removed from the spacetime. Therefore,
gravitational wave emission is calculated in a post-processing step
using the quadrupole formula. A more detailed description of the wave
extraction methods is given in the accompanying paper
\citep{dimmelmeier_02_b}.

%%%%%%%%%%%%%%%%%%%%%%%%%%%%%%%%%%%%%%%%%%%%%%%%%%%%%%%%%%%%%%%%%%%%%%%%%%%%%%%%%%%
%%%%%%%%%%%%%%%%%%%%%%%%%%%%%%%%%%%%%%%%%%%%%%%%%%%%%%%%%%%%%%%%%%%%%%%%%%%%%%%%%%%

\subsection{General relativistic hydrodynamic equations}
\label{subsec:hydro_equations}

The hydrodynamic evolution of a relativistic perfect fluid with
rest-mass current $ J^{\mu} = \rho u^{\mu} $ and energy-momentum
tensor $ T^{\mu\nu} = \rho h u^\mu u^\nu + P g^{\mu\nu} $ in a
(dynamic) spacetime $ g^{\mu\nu}$ is determined by a system of local
conservation equations, which read:
\begin{equation}
  \nabla_{\mu} J^{\mu} = 0, \qquad \nabla_{\mu} T^{\mu \nu} = 0.
  \label{eq:gr_equations_of_motion}
\end{equation}
Here $ \nabla_{\mu} $ is the covariant derivative and $ u^\mu $ is the
four-velocity of the fluid. Its three-velocity, as measured by an
Eulerian observer at rest in a spacelike hypersurface $ \Sigma_{t} $ is
\begin{equation}
  v^i = \frac{u^i}{\alpha u^t} + \frac{\beta^i}{\alpha}.
  \label{eq:three_velocity}
\end{equation}

Following the work of \citet{banyuls_97_a} we now introduce the
following set of conserved variables in terms of the primitive
(physical) hydrodynamic variables $ (\rho, v_i, \epsilon) $, with
$ \rho $ and $ \epsilon $ being the rest-mass density and specific
internal energy density, respectively:
\begin{equationarray}
  D & = & J^\mu n_\mu = \rho W,
  \label{eq:conserved_quantities_d} \\
  S^i  & =  &- \bot^i_\nu T^{\mu \nu} n_\mu = \rho h W^2 v^i,
  \label{eq:conserved_quantities_s} \\
  \tau & = & T^{\mu \nu} n_\mu n_\nu - J^\mu n_\mu = \rho h W^2 - P - D.
  \label{eq:conserved_quantities_tau}
\end{equationarray}%
In the above expressions, $ n_\mu $ is the unitary vector normal to
the slice and $ \bot^i_\nu $ denotes the projection
operator. Furthermore $ W $ is the Lorentz factor defined as
$ W = \alpha u^t $ and satisfying the relation
$ W = 1/\sqrt{1 - v_i  v^i} $. In addition,
$ h = 1 + \epsilon + P / \rho $ is the specific enthalpy and
$ P $ is the pressure.

The local conservation laws, Eq.~(\ref{eq:gr_equations_of_motion}),
can be written as a first-order, flux-conservative hyperbolic system of
equations,
\begin{equation}
  \frac{1}{\sqrt{- g}} \left[
  \frac{\partial \sqrt{\gamma} \mb{U}}{\partial x^0} +
  \frac{\partial \sqrt{- g} \mb{F}^i}{\partial x^i} \right] = \mb{Q}.
  \label{eq:hydro_conservation_equation}
\end{equation}
with the state vector, flux vector and source vector given by
\begin{equationarray}
  \mb{U} & = & [D, S_j, \tau] \\
  \mb{F}^i & = & \left[
  D \left( v^i - \frac{\beta^i}{\alpha} \right),
  S_j \left( v^i - \frac{\beta^i}{\alpha} \right) + \delta^i_j P, \right.
  \nonumber \\ 
  & & \left.
  \tau \left( v^i - \frac{\beta^i}{\alpha} \right) + P v^i \right], \\
  \mb{Q} & = & \left[ 0, T^{\mu \nu} \left(
  \frac{\partial g_{\nu j}}{\partial x^\mu} -
   {\it \Gamma}^\lambda_{\mu \nu} g_{\lambda j} \right), \right.
  \nonumber \\
  & & \left. \alpha \left( T^{\mu 0}
  \frac{\partial \ln \alpha}{\partial x^\mu} -
  T^{\mu \nu} {\it \Gamma}^0_{\mu \nu} \right) \right].
  \label{eq:hydro_conservation_equation_vectors}
\end{equationarray}%
Here $ \sqrt{-g} = \alpha \sqrt{\gamma} $, with
$ g = \det (g_{\mu \nu}) $ and $ \gamma = \det (\gamma_{ij}) $ being
the determinant of the four-metric and three-metric, respectively. In
addition, $ {\it \Gamma}^\lambda_{\mu \nu} $ are the Christoffel
symbols.

%%%%%%%%%%%%%%%%%%%%%%%%%%%%%%%%%%%%%%%%%%%%%%%%%%%%%%%%%%%%%%%%%%%%%%%%%%%%%%%%%%%
%%%%%%%%%%%%%%%%%%%%%%%%%%%%%%%%%%%%%%%%%%%%%%%%%%%%%%%%%%%%%%%%%%%%%%%%%%%%%%%%%%%
%%%%%%%%%%%%%%%%%%%%%%%%%%%%%%%%%%%%%%%%%%%%%%%%%%%%%%%%%%%%%%%%%%%%%%%%%%%%%%%%%%%

\section{Initial models for rotational core collapse}
\label{sec:initial_models}

%%%%%%%%%%%%%%%%%%%%%%%%%%%%%%%%%%%%%%%%%%%%%%%%%%%%%%%%%%%%%%%%%%%%%%%%%%%%%%%%%%%
%%%%%%%%%%%%%%%%%%%%%%%%%%%%%%%%%%%%%%%%%%%%%%%%%%%%%%%%%%%%%%%%%%%%%%%%%%%%%%%%%%%

\subsection{Equation of state}
\label{subsec:eos}

For a rotating iron core before collapse the polytropic relation
between the pressure $ P $ and the rest mass density $ \rho $,
\begin{equation}
  P = K \rho^{\gamma},
  \label{eq:polytropic_relation}
\end{equation}
with $ \gamma = \gamma_{\rm ini} = 4 / 3 $ and
$ K \sim 4.897 \times 10^{14} $ (in cgs units) is a fair approximation of
the density and pressure stratification \citep{zwerger_97_a}.

To model the physical processes which lead to the onset of collapse,
we reduce the effective adiabatic index from $ \gamma_{\rm ini} $ to
$ \gamma_1 $ on the initial time slice. During the infall phase of core
collapse, we also assume the matter of the core to obey a polytropic
EoS, Eq.~(\ref{eq:polytropic_relation}), which is consistent with the
ideal gas EoS for a compressible inviscid fluid,
\begin{equation}
  P = (\gamma - 1) \rho \epsilon,
  \label{eq:ideal_gas_eos}
\end{equation}
where $ \epsilon $ is the specific internal energy.

During and after core bounce, when the shock starts to propagate out,
the matter in the post-shock region is heated, i.e.\ kinetic energy is
dissipated into internal energy. Therefore, in this stage we demand
that the total pressure consists of a polytropic and a thermal part,
$ P = P_{\rm p} + P_{\rm th} $. We further assume that the polytropic
index jumps from $ \gamma_1 $ to $ \gamma_2 $ for densities larger
than nuclear matter density $ \rho_{\rm nuc} $. This approximates the
stiffening of the EoS. Requiring that the pressure and internal energy
are continuous at the transition density $ \rho_{\rm nuc} $, one finds
for the polytropic internal energy
\begin{equation}
  \epsilon_{\rm p} = \left\{
  \begin{array}{ll}
    \displaystyle \frac{K}{\gamma_1 - 1} \rho^{\gamma_1 - 1} \\
    \displaystyle \frac{K}{\gamma_2 - 1} \rho^{\gamma_2 - 1}
    \rho_{\rm nuc}^{\gamma_1 - \gamma_2} +
    \frac{(\gamma_2 - \gamma_1) K}{(\gamma_2 - 1) (\gamma_1 - 1)}
    \rho_{\rm nuc}^{\gamma_1 - 1} 
  \end{array}
  \right.
  \label{eq:polytropic_eps}
\end{equation}
for $ \rho \le \rho_{\rm nuc} $ and $ \rho > \rho_{\rm nuc} $,
respectively (see \citet{zwerger_95_a}). The thermal part of the
pressure, which is due to shock heating, is given by
$ P_{\rm th} = (\gamma_{\rm th} - 1) \rho \epsilon_{\rm th} $, where
$ \gamma_{\rm th} = 1.5 $ and
$ \epsilon_{\rm th} = \epsilon - \epsilon_{\rm p} $. This describes a
mixture of relativistic ($ \gamma = 4/3 $) and nonrelativistic
($ \gamma = 5/3 $) components of an ideal fluid.

From these considerations one can construct an EoS for which both the
total pressure $ P $ and the individual contributions $ P_{\rm p} $
and $ P_{\rm th} $ are continuous at $ \rho_{\rm nuc} $:
\begin{equationarray}
  P & = & \frac{\gamma - \gamma_{\rm th}}{\gamma - 1}
  K \rho_{\rm nuc}^{\gamma_1 - \gamma}
  \rho^{\gamma} - \frac{(\gamma_{\rm th} - 1) (\gamma - \gamma_1)}
  {(\gamma_1 - 1) (\gamma_2 - 1)}
  K \rho_{\rm nuc}^{\gamma_1 - 1} \rho \nonumber \\
  & & + (\gamma_{\rm th} - 1) \rho \epsilon.
  \label{eq:hybrid_eos}
\end{equationarray}%

%%%%%%%%%%%%%%%%%%%%%%%%%%%%%%%%%%%%%%%%%%%%%%%%%%%%%%%%%%%%%%%%%%%%%%%%%%%%%%%%%%%
%%%%%%%%%%%%%%%%%%%%%%%%%%%%%%%%%%%%%%%%%%%%%%%%%%%%%%%%%%%%%%%%%%%%%%%%%%%%%%%%%%%

\subsection{Rotating relativistic stars in equilibrium}
\label{subsec:rns_construction}

Our initial models are obtained using Hachisu's self-consistent field
method \citep{komatsu_89_a} which is a general relativistic
method for solving the hydrostatic equilibrium equations for rotating
polytropic matter distributions. The general metric to describe
rotating axisymmetric relativistic matter configurations in
equilibrium is
\begin{equationarray}
  ds^2 & = & - e^{2 \hat{\nu}} dt^2 + e^{2 \hat{\alpha}} (dr^2 + r^2 d\theta^2) 
  \nonumber \\
  & & + e^{2 \hat{\beta}} r^2 \sin^2 \theta (d\varphi - \omega dt)^2,
  \label{eq:rrs_metric}
\end{equationarray}%
with the metric potentials $ \hat{\nu} $, $ \hat{\alpha} $,
$ \hat{\beta} $ and $ \omega $.

The only nonvanishing three-velocity component is $ v^3 $, the
velocity of the fluid as measured by a ZAMO (zero angular momentum
observer). Thus, the Lorentz factor is given by
$ W = 1 / \sqrt{1 - v_3 v^3} $. With the definitions
$ u^0 = W e^{- \hat{\nu}} $, and $ u^3 = \Omega W e^{- \hat{\nu}} $,
where $ \Omega $ is the angular velocity of the fluid as measured from
infinity, the four-velocity reads:
\begin{equation}
  u^\mu = W e^{- \hat{\nu}} (1, 0, 0, \Omega).
  \label{eq:rrs_four_velocity}
\end{equation}
Thus, one gets for the $ \varphi $-component of the three-velocity:
\begin{equationarray}
  v^3 & = & \frac{u^3}{u^0 e^{\hat{\nu}}} -
  \frac{\omega}{e^{\hat{\nu}}} = e^{- \hat{\nu}} (\Omega - \omega),
  \label{eq:rrs_phi_3_velocity_upper} \\
  v_3 & = & e^{2 \hat{\beta} - \hat{\nu}} r^2 \sin^2 \theta
  (\Omega - \omega),
  \label{eq:rrs_phi_3_velocity_lower} \\
  v_\varphi & \equiv & \sqrt{v_3 v^3} = e^{\hat{\beta} -
  \hat{\nu}} r \sin \theta (\Omega - \omega).
  \label{eq:rrs_phi_3_velocity}
\end{equationarray}%
Here $ v_\varphi $ is the proper rotation velocity with respect to the
ZAMO. The specific angular momentum of the fluid $ j \equiv u^0 u_3 $
is given by
\begin{equation}
  j = W^2 e^{2 (\hat{\beta} - \hat{\nu})} r^2 \sin^2 \theta
  (\Omega - \omega) = \frac{v_3 v^3}{(1 - v_3 v^3) (\Omega - \omega)}.
  \label{eq:rrs_j}
\end{equation}

Using the Einstein equations one can derive the equation of hydrostatic
equilibrium for an axisymmetric mass distribution rotating with the
angular velocity $ \Omega = \Omega(r) $:
\begin{equation}
  \frac{\nabla P}{\rho h} + \nabla \hat{\nu} -
  \frac{v_\varphi}{1 - v_\varphi^2} \nabla v_\varphi + j \nabla \Omega = 0.
  \label{eq:diff_equilibrium_gr}
\end{equation}

For barotropes where $ p $ depends only on $ \rho $ the integrability
of Eq.~(\ref{eq:diff_equilibrium_gr}) requires that $ j $ is a
function of $ \Omega $, only. Hence, the simplest choice for a
rotation law is
\begin{equation}
  j = A^2 (\Omega_{\rm c} - \Omega),
  \label{eq:rotation_law_gr}
\end{equation}
where $ \Omega_{\rm c} $ is the value of $ \Omega $ at the coordinate
center and $ A $ is a positive constant \citep{komatsu_89_a}. In the
Newtonian limit the rotation law~(\ref{eq:rotation_law_gr}) reduces to
\begin{equation}
  \frac{\Omega}{\Omega_{\rm c}} = \frac{A^2}{A^2 + d^2} 
                                =  \left\{
  \begin{array}{lll}
    1                             \quad & \mbox{for} \,\, 
                                  A \rightarrow \infty \\
    \displaystyle \frac{A^2}{d^2} \quad & \mbox{for} \,\, 
                                  A \rightarrow 0 
  \end{array}
  \right.
  \label{eq:rotation_law_newtonian}
\end{equation}
where $ d = r \sin \theta $ is the distance from the rotation axis.
The upper case in Eq.~(\ref{eq:rotation_law_newtonian}) corresponds to
a rigid rotator and the lower one to a configuration with constant
specific angular momentum.

Using the rotation law specified in Eq.~(\ref{eq:rotation_law_gr})
one can integrate the equation of hydrostatic
equilibrium~(\ref{eq:diff_equilibrium_gr}) to obtain
\citep{komatsu_89_a}
\begin{equation}
  \ln h + \hat{\nu} + \frac{1}{2} \ln (1 - v_{\varphi}^2) - \frac{1}{2} A^2
  (\Omega - \Omega_{\rm c})^2 = {\rm const},
  \label{eq:int_equilibrium_gr}
\end{equation}
The four-metric components $ \hat{\alpha} $, $ \gamma = \hat{\beta} +
\hat{\nu} $, $ \delta = \hat{\nu} - \hat{\beta} $ and $ \omega $ are
given by the four coupled partial differential equations
\citep{komatsu_89_a}
\begin{equationarray}
  \Delta \left( \delta e^{\gamma / 2} \right) & = & S_\delta, \\
  \label{hscf_metric_equations_1}
  \left( \Delta + \frac{1}{r} \partial_r - \frac{\mu}{r^2}
  \partial_\mu \right) \gamma e^{\gamma/2} & = & S_\gamma, \\
  \label{hscf_metric_equations_2}
  \left( \Delta + \frac{2}{r} \partial_r - \frac{2\mu}{r^2}
  \partial_\mu \right) \omega e^{(\gamma - 2 \delta)/ 2 } & = &
  S_\omega, \\
  \label{hscf_metric_equations_3}
  \partial_\mu \hat{\alpha} & = & S_{\hat{\alpha}},
  \label{hscf_metric_equations_4}
\end{equationarray}%
where $ \Delta $ is the Laplacian and $ \mu \equiv \cos \theta $. The
source terms $ S_\delta $, $ S_\gamma $, $ S_\omega $, and
$ S_ {\hat{\alpha}} $ depend in general on all four metric functions.
In order to solve this system of equations the first three PDEs are
converted into integral equations using Green functions which in turn
are expanded into Legendre polynomials. In discretized form, these
equations are iterated until convergence is obtained. During each
iteration, the discretized equation~(\ref{hscf_metric_equations_4})
for $ \hat{\alpha} $ is integrated.

%%%%%%%%%%%%%%%%%%%%%%%%%%%%%%%%%%%%%%%%%%%%%%%%%%%%%%%%%%%%%%%%%%%%%%%%%%%%%%%%%%%
%%%%%%%%%%%%%%%%%%%%%%%%%%%%%%%%%%%%%%%%%%%%%%%%%%%%%%%%%%%%%%%%%%%%%%%%%%%%%%%%%%%

\subsection{Initial models and collapse parameters}
\label{subsec:initial_models}

\begin{table}
  \centering
  \caption{Set of simulated collapse models: $ A $ is the degree of 
    differential rotation, $ \beta_{\rm rot\, ini} $ is the initial rotation
    rate (i.e., the initial ratio of rotational energy and the absolute 
    value of the gravitational binding energy), and $ \gamma_1 $
    ($ \gamma_2 $) is the adiabatic index at densities below (above)
    nuclear density. See text for further details.}
  \begin{tabular}{lcccc} 
    \hline  
    \multicolumn{1}{c}{Collapse~~~~~} & 
    $ A $ & $ \beta_{\rm rot\, ini} $ & $ \gamma_1 $ & $ \gamma_2 $ \\ 
    \multicolumn{1}{c}{model~~~~~} & 
    [$ 10^8 {\rm cm} $] & [\%] \\ 
    \hline 
    A1B1G1               &   50.0 &   0.25 & 1.325 & 2.5 \\ 
    A1B2G1               &   50.0 & \wz0.5 & 1.325 & 2.5 \\ 
    A1B3G1               &   50.0 & \wz0.9 & 1.325 & 2.5 \\
    A1B3G2               &   50.0 & \wz0.9 & 1.320 & 2.5 \\
    A1B3G3               &   50.0 & \wz0.9 & 1.310 & 2.5 \\
    A1B3G5               &   50.0 & \wz0.9 & 1.280 & 2.5 \\ [1.5 ex]
    A2B4G1               & \wz1.0 & \wz1.8 & 1.325 & 2.5 \\ [1.5 ex]
    A3B1G1               & \wz0.5 &   0.25 & 1.325 & 2.5 \\
    A3B2G1               & \wz0.5 & \wz0.5 & 1.325 & 2.5 \\
    A3B2G2               & \wz0.5 & \wz0.5 & 1.320 & 2.5 \\
    A3B2G4$_{\rm soft} $ & \wz0.5 & \wz0.5 & 1.300 & 2.0 \\
    A3B2G4               & \wz0.5 & \wz0.5 & 1.300 & 2.5 \\
    A3B3G1               & \wz0.5 & \wz0.9 & 1.325 & 2.5 \\
    A3B3G2               & \wz0.5 & \wz0.9 & 1.320 & 2.5 \\
    A3B3G3               & \wz0.5 & \wz0.9 & 1.310 & 2.5 \\
    A3B3G5               & \wz0.5 & \wz0.9 & 1.280 & 2.5 \\
    A3B4G2               & \wz0.5 & \wz1.8 & 1.320 & 2.5 \\
    A3B5G4               & \wz0.5 & \wz4.0 & 1.300 & 2.5 \\ [1.5 ex]
    A4B1G1               & \wz0.1 &   0.25 & 1.325 & 2.5 \\
    A4B1G2               & \wz0.1 &   0.25 & 1.320 & 2.5 \\
    A4B2G2               & \wz0.1 & \wz0.5 & 1.320 & 2.5 \\
    A4B2G3               & \wz0.1 & \wz0.5 & 1.310 & 2.5 \\
    A4B4G4               & \wz0.1 & \wz1.8 & 1.300 & 2.5 \\
    A4B4G5               & \wz0.1 & \wz1.8 & 1.280 & 2.5 \\
    A4B5G4               & \wz0.1 & \wz4.0 & 1.300 & 2.5 \\
    A4B5G5               & \wz0.1 & \wz4.0 & 1.280 & 2.5 \\
    \hline
  \end{tabular}
  \label{tab:collapse_models}
\end{table}

For a given adiabatic index $ \gamma_{\rm rot\, ini} $, the initial
models are determined by three parameters: The central density
$ \rho_{\rm c\,ini} $, the degree of differential rotation $ A $ (see
Eq.~(\ref{eq:rotation_law_gr})), and the rotation rate
$ \beta_{\rm rot\, ini} $, which is given by the ratio of rotational
energy and the absolute value of the gravitational binding energy. The
central density of an iron core is about $ 10^{10} {\rm\ g\ cm}^{-3} $.
We adopt this value for $ \rho_{\rm c\,ini} $ in all initial models.

Due to the lack of consistent models of rotating iron cores from
stellar evolution calculations, we follow \citet{zwerger_97_a} and
treat $ A $ and $ \beta_{\rm rot\, ini} $ as free parameters of the
initial models. The only other free parameters are $ \gamma_1 $, the
adiabatic index at subnuclear densities ($ \rho < \rho_{\rm nuc} $),
and $ \gamma_2 $, the adiabatic index at supranuclear densities
($ \rho \ge \rho_{\rm nuc} $). As in \citet{zwerger_97_a}, we set
$ \gamma_2 = 2.5 $ and $ \gamma_1 $ equal to 1.325, 1.320, 1.310, 1.300
or 1.280, respectively. In total, we have evolved 26 models (see
Table~\ref{tab:collapse_models}). One of the models
(A3B2G4$ _{\rm soft} $) has also been run with $ \gamma_2 = 2.0 $ in
order to test the influence of a softer supranuclear EoS on the
collapse dynamics. The nuclear density is set to
$ \rho_{\rm nuc} = 2.0 \times 10^{14} {\rm\ g\ cm}^{-3} $ in all
simulations.

%%%%%%%%%%%%%%%%%%%%%%%%%%%%%%%%%%%%%%%%%%%%%%%%%%%%%%%%%%%%%%%%%%%%%%%%%%%%%%%%%%%
%%%%%%%%%%%%%%%%%%%%%%%%%%%%%%%%%%%%%%%%%%%%%%%%%%%%%%%%%%%%%%%%%%%%%%%%%%%%%%%%%%%

\subsection{Atmosphere treatment}
\label{subsec:atmosphere}

One common problem of many hydrodynamic codes is their inability to
handle regions where the density is zero (or very small compared to
the typical densities of interest). Such regions are encountered, for
example, when simulating pulsating stars on an Eulerian grid, or when
simulating rotating stars on a spherical grid. In the first case a
very low density ``atmosphere'' is required outside the stellar surface
into which the star can expand during its pulsations. In the second
case the stellar surface is flattened due to centrifugal forces,
i.e.\ part of the computational grid extends outside the flattened
surface and must be filled with some low density atmosphere. Thus,
when computing initial models of rotating cores we introduce a low
density atmosphere outside the stellar matter distribution (see
Table~\ref{tab:atmosphere_model}). An atmosphere is assumed in all
zones where the density $ \rho $ is below a prescribed threshold value
which is practically defined as a fraction $ f_{\rm atm\,thr} $ of the
central density of the initial model, i.e.\ the density of the
atmosphere $ \rho_{\rm atm} = f_{\rm atm} \rho_{\rm c\,ini} $. During
a simulation the atmosphere is reset after each timestep. This ensures
that it adapts to the time-dependent ``shape'' of the stellar surface.

\begin{table}
  \centering
  \caption{Atmosphere parameters used in our core collapse
    simulations. For the extremely differentially rotating
    models A4 the threshold value $ f_{\rm atm\,thr} $ is 
    in the range $ 5 \times 10^{- 4} $ to $ 1 \times 10^{- 4} $.}
  \begin{tabular}{ccccc}
    \hline
    $ f_{\rm atm\,thr} $ &
    $ f_{\rm atm} $ & $ \rho_{\rm atm} $ & EoS & $ v_i $ \\
    & & [g cm$^{-1}$] & & \\
    \hline
    $ 10^{- 5} $ & $ 10^{- 10} $ & 1 &
    $ P = K \rho_{\rm atm}^{\gamma_1} $ & 0 $ \rule{0 em}{1 em} $ \\
    \hline
  \end{tabular}
  \label{tab:atmosphere_model}
\end{table}

For the values specified in Table~\ref{tab:atmosphere_model} and an
initial central density
$ \rho_{\rm c\,ini} = 10^{10} \rm{\ g\ cm}^{-3} $ the homogeneous
atmosphere has a density $ \rho_{\rm atm} = 1 \rm{\ g\ cm}^{-3} $. For
all practical purposes this can be considered as a low density
atmosphere, which has no influence on the dynamics of core collapse,
the only tradeoff being a slight loss of angular momentum to the
atmosphere from regions of the star which are close to the boundary
(see Section~\ref{subsec:conservation}). The EoS in the atmosphere is
assumed to be polytropic, too.

%%%%%%%%%%%%%%%%%%%%%%%%%%%%%%%%%%%%%%%%%%%%%%%%%%%%%%%%%%%%%%%%%%%%%%%%%%%%%%%%%%%
%%%%%%%%%%%%%%%%%%%%%%%%%%%%%%%%%%%%%%%%%%%%%%%%%%%%%%%%%%%%%%%%%%%%%%%%%%%%%%%%%%%
%%%%%%%%%%%%%%%%%%%%%%%%%%%%%%%%%%%%%%%%%%%%%%%%%%%%%%%%%%%%%%%%%%%%%%%%%%%%%%%%%%%

\section{Numerical implementation}
\label{sec:numerical_implementation}

%%%%%%%%%%%%%%%%%%%%%%%%%%%%%%%%%%%%%%%%%%%%%%%%%%%%%%%%%%%%%%%%%%%%%%%%%%%%%%%%%%%
%%%%%%%%%%%%%%%%%%%%%%%%%%%%%%%%%%%%%%%%%%%%%%%%%%%%%%%%%%%%%%%%%%%%%%%%%%%%%%%%%%%

\subsection{Computational grid}
\label{subsec:grid}

We use spherical polar coordinates $ (t, r, \theta, \varphi) $, and
assume axial symmetry with respect to the rotation axis, i.e.\ none of
the variables depends on the azimuthal coordinate $ \varphi $.
However, motions in the $ \varphi $-direction are allowed, and the
associated velocity $ v_3 $ and momentum $ S_3 $ are nonzero.

We further assume symmetry with respect to the equator at
$ \theta = \pi / 2 $. Therefore, $ \theta $ spans the interval between
0 and $ \pi / 2 $. The angular grid consisting of $ n_\theta $ zones is
equally spaced, with $ \theta_j $ denoting the coordinates of the cell
centers in angular direction ($ 1 \le j \le n_\theta $). The $ n_r $
radial zones (zone centers located at $ r_i $ ) are logarithmically
spaced. Each computational cell $ (r_i, \theta_j) $ is bounded by
interfaces at $ r_{\im} $, $ r_{\ip} $, $ \theta_{\jm} $, and
$ \theta_{\jp} $, respectively. The angular size of a zone is
$ \Delta \theta = {1\over2} \pi / n_\theta $, while the radial size is
given by $ \Delta r_i = r_{\ip} - r_{\im} $. The radial size of the
three innermost cells is enlarged in order to ease the timestep
restriction; for $ i > 3 $,
$ \Delta r_{i+1} / \Delta r_i = {\rm const} $. Note that
$ r_{1\over2} = 0 $ is the origin, and $ r_{n_r + {1\over2}} = R $
defines the outer boundary of the grid. The rotation axis coincides
with $ \theta_{1\over2} = 0 $, and
$ \theta_{n_\theta + {1\over2}} = {1\over2} \pi $ denotes the
equatorial plane. We use $ n_r = 200 $ radial zones and
$ n_\theta = 30 $ angular zones, which corresponds to an angular
resolution of $ 3^\circ $. The radial size $ \Delta r_i $ of the
innermost cells is about 500~m. Convergence tests show that this grid
resolution is sufficient to resolve the important flow features of
core collapse dynamics.

%%%%%%%%%%%%%%%%%%%%%%%%%%%%%%%%%%%%%%%%%%%%%%%%%%%%%%%%%%%%%%%%%%%%%%%%%%%%%%%%%%%
%%%%%%%%%%%%%%%%%%%%%%%%%%%%%%%%%%%%%%%%%%%%%%%%%%%%%%%%%%%%%%%%%%%%%%%%%%%%%%%%%%%

\subsection{Boundary conditions}
\label{subsec:boundary_conditions}

We impose symmetry conditions for both hydrodynamic and metric
quantities at the center ($ r = 0 $) and the rotation axis
($ \theta = 0 $), and at the equatorial plane ($ \theta = \pi / 2 $).
Quantities are assumed to be either symmetric or antisymmetric across
a boundary. We combine these two options in the following notation for
a quantity $ q_{i,j} $:
\begin{equationarray}
  q_{-{1\over2},j}         & = & \pm \, q_{{1\over2},j}, \\
  q_{i,n_\theta+{1\over2}} & = & \pm \, q_{i,n_\theta-{1\over2}}, \\
  q_{i,-{1\over2}}         & = & \pm \, q_{i,{1\over2}},
  \label{eq:symmetry_condition_notation_1}
\end{equationarray}%
where
\begin{equation}
  \pm = \left\{
  \begin{array}{l}
    + \qquad \mbox{symmetric}, \\
    - \qquad \mbox{antisymmetric}.
  \end{array}
  \right.
  \label{eq:symmetry_condition_notation_2}
\end{equation}

The symmetry conditions for the scalar quantities $ \rho $,
$ \epsilon $, $ \phi $ and $ \alpha $ are straightforward. Like their
Newtonian counterparts $ \rho $, $ \epsilon $ and $ \Phi $ they have
to be continuous across all boundaries. Thus, the scalar quantities
are assumed symmetric (+) with respect to all boundaries. Contrary to
Newtonian hydrodynamics, the symmetry conditions for the vector
quantities are nontrivial, as they occur both in covariant and
contravariant form.

By defining ``physical'' velocities (the same can be done for the shift 
vector components $ \beta^i $),
\begin{equationarray}
  v_r & \equiv & \sqrt{v_1 v^1} =
  \phi^{-2} v_1 = \phi^2 v^1,
  \label{eq:physical_velocities_1} \\
  v_\theta & \equiv & \sqrt{v_2 v^2} =
  \phi^{-2} r^{-1} v_2 = \phi^2 r v^2,
  \label{eq:physical_velocities_2} \\
  v_\varphi & \equiv & \sqrt{v_3 v^3} =
  \phi^{-2} r^{-1} \sin^{-1} \theta v_3 =
  \phi^2 r \sin \theta v^3,
  \label{eq:physical_velocities_3}
\end{equationarray}%
which are bounded by the speed of light
($ 0 \le |v_r|, |v_\theta|, |v_\varphi| \le 1 $), we can specify the
symmetry conditions from Newtonian hydrodynamics. Keeping in mind the
symmetry behaviour of the geometrical terms in the three-metric
components $ \gamma_{ij} $ and $ \gamma^{ij} $, the symmetry
conditions for the ``physical velocity vector'' and the shift vector
read:
\begin{displaymath}
  \begin{tabular}{l|ccc|ccc}
    & $ v_r $ & $ v_\theta $ & $ v_\varphi $ & $ \beta^1 $ & $ \beta^2 $ &
    $ \beta^3 $ \\
    \hline
    center  & $-$ & $-$ & $-$ & $-$ & $+$ & $+$ \\
    pole    & $+$ & $-$ & $-$ & $+$ & $-$ & $+$ \\
    equator & $+$ & $-$ & $+$ & $+$ & $-$ & $+$ \\
  \end{tabular}
\end{displaymath}%
Here we demand that (i) all velocities have to vanish at the center
(no mass flow across the center), that (ii) the meridional velocity
$ v_\theta $ is zero along the rotation axis and in the equatorial plane
(no mass flow across these boundaries), and that (iii) the azimuthal
velocity $ v_\varphi $ is zero along the rotation axis. Hence, these
entries are set to ($-$), while all others, where the velocities are
continuous, are set to (+). The shift vector components $ \beta^i $
are treated according to the contravariant three-velocity vector
$ v^i $, as the shift vector corresponds to a ``coordinate velocity''.

At the outer radial boundary, the hydrodynamic quantities are
extrapolated from the interior to the boundary zones. However, this
procedure cannot be used for the metric quantities, as they are
determined by elliptic equations which define a boundary value
problem. We thus determine their boundary conditions by matching the
interior metric to an exterior spacetime metric. We assume that the
exterior spacetime is given by the Schwarzschild solution for a
spherically symmetric vacuum spacetime. The quality of this
approximation is sufficiently good for all practical purposes provided
the spacetime at the surface of the star does not deviate too much
from spherical symmetry. This is true for iron core initial data, and
also holds for rotational core collapse, as the outer boundary of the
radial grid is fixed (at the equatorial radius of the initial model)
at about $ 1500 {\rm\ km} $ and as the monopole term always dominates
the gravitational field.

The Schwarzschild solution in isotropic coordinates is
\begin{equationarray}
  ds^2 & = & - \frac{\left( 1 - \frac{M_{\rm grav}}{2 r} \right)^2}
  {\left( 1 + \frac{M_{\rm grav}}{2 r} \right)^2} \, dt^2 \nonumber \\
  & & + \left(1 + \frac{M_{\rm grav}}{2 r} \right)^4 (dr^2 + r^2 d\Omega^2).
  \label{eq:isotropic_schwarzschild_metric}
\end{equationarray}%
For a vanishing shift vector this metric can be exactly matched to
the conformally flat interior metric:
\begin{equation}
  \phi = 1 + \frac{M_{\rm grav}}{2 r}, \qquad
  \alpha = \frac{\left( 1 - \frac{M_{\rm grav}}{2 r} \right)}
  {\left( 1 + \frac{M_{\rm grav}}{2 r} \right)}, \qquad
  \beta^i = 0.
  \label{eq:vacuum_matching}
\end{equation}
As the gravitational mass $ M_{\rm grav} $ is conserved and as
$ \beta^i $ is always close to zero at the outer boundary, this matching
to the isotropic Schwarzschild metric provides an acceptable outer
boundary condition for the metric coefficients $ \phi $, $ \alpha $
and $ \beta^i $.

We stress that these boundary conditions are only an approximation to
the exact vacuum spacetime solution, since no rotating spacetime can
be described by a conformally flat metric. By the above matching we
neglect effects like frame dragging ($ \beta^3 \ne 0 $), which would
in general be present in the vacuum spacetime region outside the
computational domain.

%%%%%%%%%%%%%%%%%%%%%%%%%%%%%%%%%%%%%%%%%%%%%%%%%%%%%%%%%%%%%%%%%%%%%%%%%%%%%%%%%%%
%%%%%%%%%%%%%%%%%%%%%%%%%%%%%%%%%%%%%%%%%%%%%%%%%%%%%%%%%%%%%%%%%%%%%%%%%%%%%%%%%%%

\subsection{The hydrodynamic solver}
\label{subsec:hydro_solver}

The hydrodynamic solver of our code performs the numerical integration
of system~(\ref{eq:hydro_conservation_equation}) using a so-called
Godunov-type (high-resolution shock-capturing -- HRSC) scheme. Such schemes
are specifically designed to solve nonlinear hyperbolic systems of
conservation laws (see, e.g., \citet{toro_97_a} for definitions). In a
Godunov-type method the knowledge of the characteristic structure of
the equations is essential to design a solution procedure based upon
either exact or approximate Riemann solvers. These solvers compute,
at every cell-interface of the numerical grid, the solution of local
Riemann problems. Therefore, they automatically guarantee the proper
capturing of discontinuities which may arise naturally in the solution
space of a nonlinear hyperbolic system.

The time update of system~(\ref{eq:hydro_conservation_equation}) from
$ t^n $ to $ t^{n + 1} $ is performed according to the following
conservative algorithm:
\begin{equationarray}
  \mb{U}_{i,j}^{n + 1} = \mb{U}_{i,j}^{n} & - & \frac{\Delta t}{\Delta r_i}
  (\widehat{\mb{\!F}}^r_{i + 1/2,j} - \widehat{\mb{\!F}}^r_{i-1/2,j})
  \nonumber \\
  & - & \frac{\Delta t}{\Delta \theta}
  (\widehat{\mb{\!F}}^{\theta}_{i,j + 1/2} -
  \widehat{\mb{\!F}}^{\theta}_{i,j - 1/2})
  \nonumber \\
  & + & \Delta t \, \mb{Q}_{i,j}.
\end{equationarray}%
The numerical fluxes $ \widehat{\mb{\!F}}^{r} $ and
$ \widehat{\mb{\!F}}^{\theta} $ are computed by means of Marquina's
approximate flux-formula \citep{donat_96_a}. The fluxes are obtained
independently for each direction. The time update of the state-vector
$ \mb{U} $ is done simultaneously in both spatial directions using a
method of lines in combination with a second-order (in time)
conservative Runge--Kutta scheme. Moreover, in order to set up a
family of local Riemann problems at every cell-interface we use the
piecewise parabolic method (PPM) of \citet{colella_84_a} for the
reconstruction of cell interface values, which provides third-order
accuracy in space.

Time derivatives of all metric quantities ($ \phi $, $ \alpha $, and
$ \beta^i $) are required to compute the Christoffel symbols, and hence
the source term $ \mb{Q} $. As the equations for the metric in the CFC
approximation do not provide explicit expressions for the time
derivatives of these quantities, we approximate them numerically by
discretized derivatives using values from the two time slices
$ \Sigma_{t^n} $ and $ \Sigma_{t^{n-1}} $, e.g.\
\begin{equation}
  \left( \frac{\partial \phi}{\partial t} \right)_{i,j}^n \!\! =
  \frac{\phi_{i,j}^n - \phi_{i,j}^{n-1}}{\Delta t^{n-1}}.
  \label{eq:metric_time_derivatives}
\end{equation}

Finally we note that after the hydrodynamics update we need to recover
the primitive variables $(\rho,v_i, \epsilon) $ from the conserved
ones $ (D, S_i, \tau) $. Since the relation between these two sets is
only given implicitely, we have to resort to a Newton-Raphson iterative
method. This procedure is discussed in Appendix~\ref{app:recovery}.

%%%%%%%%%%%%%%%%%%%%%%%%%%%%%%%%%%%%%%%%%%%%%%%%%%%%%%%%%%%%%%%%%%%%%%%%%%%%%%%%%%%
%%%%%%%%%%%%%%%%%%%%%%%%%%%%%%%%%%%%%%%%%%%%%%%%%%%%%%%%%%%%%%%%%%%%%%%%%%%%%%%%%%%

\subsection{The metric solver}
\label{subsec:metric_solver}

One of the standard methods to solve an elliptic linear system like
Poisson's equation on a numerical grid is to discretize the equation.
Let the vector of unknowns be $ \mb{u} = (u_{i, j}) $, and the vector
of sources be $ \mb{f} = (f_{i, j}) $, where $ (i,j) $ labels the
position $ (r_i, \theta_j) $ on the grid. This discretization leads to
a linear matrix equation with the right hand side being the source
vector of the original equation:
\begin{equation}
  \mb{A} \mb{u} = \mb{f}.
  \label{eq:general_linear_problem}
\end{equation}
In general the matrix $ \mb{A} $ is sparse, i.e.\ the number of
nonzero elements is much smaller than the total number of
elements. For solving sparse linear systems there exist a variety of
efficient numerical methods and software packages for various computer
architectures.

Although the CFC metric
equations~(\ref{eq:metric_equation_1}--\ref{eq:metric_equation_3}) do
not form a linear system like Poisson's equation, they can still be
reduced to a linear system. Our strategy is as follows: We first
define a vector of unknowns, whose components are the five metric
quantities:
\begin{equation}
  \hat{u} = (u^k) = (\phi, \alpha \phi, \beta^1, \beta^2, \beta^3).
  \label{eq:metric_solution_vector}
\end{equation}
Then the metric equations can be written as
\begin{equation}
  \hat{f} (\hat{u}) = 0,
  \label{eq:metric_equations_vector}
\end{equation}
with $ \hat{f} $ denoting the vector of the metric equations for
$ \hat{u} $.

The explicit form of the metric
equations~(\ref{eq:metric_equations_vector}) is given in
Appendix~\ref{app:metric_equations}. We discretize these equations
on the $ (r, \theta) $-grid, denoting the metric components $ u^k $
($ k = 1, \dots, 5 $) at the cell-center $ (r_i, \theta_j) $ by
$ u_{i, j}^k $. Their values at cell-interfaces (needed for the hydrodynamic
solver) are obtained by interpolation. The radial and angular
derivatives are approximated by central finite differences. A 9-point
stencil is required to formulate the discretized equations. Hence,
the system of metric equations, when discretized, gives rise to a
nonlinear system of dimension $ 5 \times n_r \times n_\theta $,
\begin{equation}
  \mb{\hat{f}} (\mb{\hat{u}}) = 0,
  \label{eq:discretized_metric_vector_equations}
\end{equation}
with the vector of discretized equations
$ \mb{\hat{f}} = \hat{f}_{i,j} = f_{i,j}^k $ for the discretized
unknowns $ \mb{\hat{u}} = \hat{u}_{l,m} = u_{l,m}^n $. For this system
we have to find the roots $ u_{i,j}^l $.

We use a multi-dimensional Newton--Raphson iteration method to solve
the nonlinear system, i.e.\ the nonlinear system of dimension
$ 5 \times n_r \times n_\theta $ is reduced to a linear one of the same
dimension for the Jacobi matrix of $ \mb{\hat{f}} $. This linear
matrix equation has to be solved once in each iteration step.
For our set of equations the Jacobi matrix of the nonlinear system is
in general both sparse and diagonally dominated, which reduces the
complexity of the linear problem significantly.

The performance of our axisymmetric general relativistic code is
mostly determined by the computation time spent in the metric solver,
and thus by the computational efficiency of the linear solver.
Therefore, the linear problem has to be solved as efficiently as
possible. After extensive investigation we have decided to use a block
tridiagonal sweeping method \citep{potter_73_a}. As the matrix which
defines the linear problem has nine bands of blocks rather than three
(due to the 9-point stencil; see above), we combine the left three,
the middle three, and the right three bands, collecting them in three
bands of submatrices -- left, diagonal and right. Although these
submatrices are sparse, the block tridiagonal sweeping algorithm leads
to dense submatrices which have to be inverted. This is achieved using
a standard $ LU $ decomposition scheme for dense matrices as
implemented in the LAPACK numerical library.

As far as computational performance is concerned, the block
tridiagonal sweeping method exhibits an $ n_r \times n_\theta^2 $
behavior, and is thus superior to both the Gauss--Jordan elimination
and the bi-conjugate gradient stabilized methods which we have also
tried. In particular, as it is a direct solver it does not suffer from
convergence problems.

The tolerance measure we use to control convergence of the
Newton--Raphson method is the maximum increment of the solution
vector on the whole computational grid:
\begin{equation}
  \Delta \hat{u}^s_{\rm max} =
  \max\,(\Delta \mb{\hat{u}}^s) =
  \max\,(\Delta u_{i,j}^{k\,s}).
  \label{eq:maximum_solution_vector}
\end{equation}
We are able to meet this convergence criterion for a tolerance level
of $ \Delta \hat{u}^s_{\rm max} \le 10^{- 15} $ in all our
computations.

\begin{figure}
  \resizebox{\hsize}{!}{\includegraphics{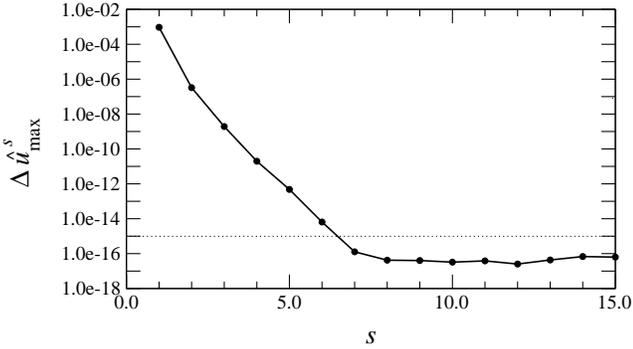}}
  \caption{Decrease of the maximum increment of the unknown vector
    $ \Delta \hat{u}^s_{\rm max} $ with the number of
    Newton--Raphson iterations $ s $ for a typical metric computation.
    Until the tolerance measure $ \Delta \hat{u}^s_{\rm max} $ reaches
    the limit set by the machine precision at $ s \approx 8 $, it
    decreases exponentially. After that it saturates well below the
    precision threshold used in the code, which is marked by the
    horizontal dotted line.}
  \label{fig:linear_solver}
\end{figure}

In Fig.~\ref{fig:linear_solver} we show how
$ \Delta \hat{u}^s_{\rm max} $ decreases
with the number of iterations $ s $ in the metric computation of a
typical core collapse model around the time of maximum density. The
convergence measures exhibits an inverse exponential behavior until
saturation is reached. This occurs when the increment of
$ \mb{\hat{u}}^s $ approaches machine precision.

In addition, we monitor the residual of the function
vector $ r_{\mbsm{\!\hat{f}}} = \mb{\hat{f}} (\mb{\hat{u}}^s)) $
on the entire computational grid. If $ \mb{\hat{u}}^s $ is the exact
solution of the discretized equations, then
$ r_{\!\mbsm{\hat{f}}} = 0 $. In practice we obtain values for
$ r_{\!\mbsm{\hat{f}}} $ which are as small as roughly
$ 10^{-9} $ of the individual terms of
$ \mb{\hat{f}} (\mb{\hat{u}}^s)) $. This measure ensures convergence
to the correct solution of the metric equations.

%%%%%%%%%%%%%%%%%%%%%%%%%%%%%%%%%%%%%%%%%%%%%%%%%%%%%%%%%%%%%%%%%%%%%%%%%%%%%%%%%%%
%%%%%%%%%%%%%%%%%%%%%%%%%%%%%%%%%%%%%%%%%%%%%%%%%%%%%%%%%%%%%%%%%%%%%%%%%%%%%%%%%%%

\subsection{Metric extrapolation}
\label{subsec:metric_extrapolation}

We have performed comparisons between the execution time for one
metric solution (for different linear solver methods) and one
hydrodynamic time step. The comparison shows that, particularly for
grid sizes appropriate for our collapse simulations, the computation
is completely dominated by the metric solver: One metric computation
can be as time-consuming as about 100 hydrodynamic time steps!

\begin{figure}
  \resizebox{\hsize}{!}{\includegraphics{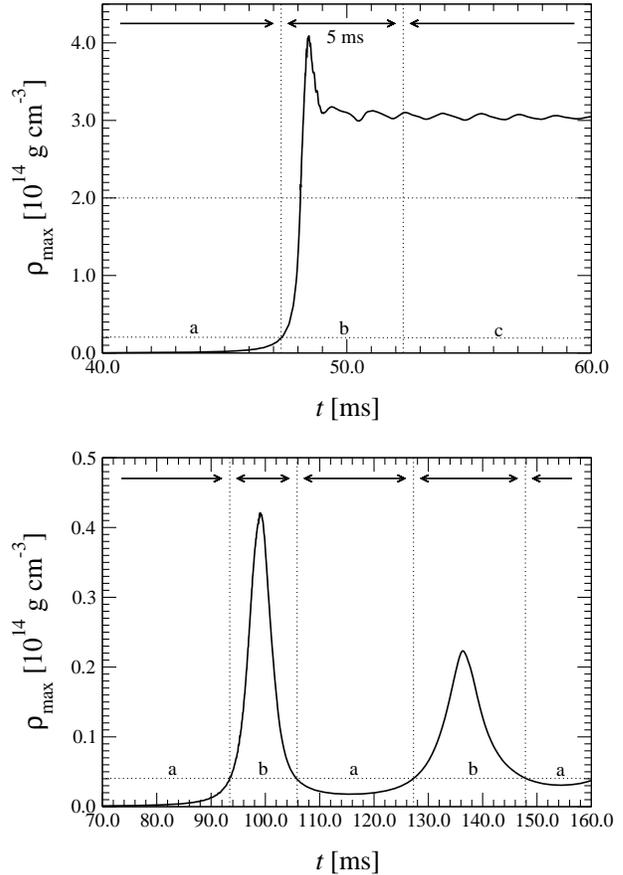}}
  \caption{The collapse phases used for specifying the metric
    resolution for regular collapse models (upper panel) and multiple
    bounce collapse models (lower panel). In a regular collapse, the
    metric is calculated on every $ n_{\rm m}^{(1)} $th time step in
    the pre-bounce phase (a). In the bounce phase (b), defined by
    $ \rho_{\rm max}>0.1 \rho_{\rm nuc}$, the metric is calculated
    every $ n_{\rm m}^{(2)} $th time step. In the post-bounce phase
    (c), which begins $ 5 {\rm\ ms} $ later, the metric is calculated
    every $ n_{\rm m}^{(3)} $th time step. In a multiple bounce model,
    the inter-bounce phases (a), during which the metric is calculated
    on every $ n_{\rm m}^{(1)} $th time step, are interspersed with
    bounce phases (b), where the metric is calculated every
    $ n_{\rm m}^{(2)} $th time step. Due to their lower average maximum
    densities, the density threshold is specified by
    $ \rho_{\rm max} \ge 0.02 \rho_{\rm nuc} $ in these models.}
  \label{fig:metric_refinement}
\end{figure}

On the other hand, as long as the metric quantities do not change too
rapidly with time, it is a fair approximation to solve for the metric
not at every time step. We introduce the metric resolution parameter
$ \Delta n_{\rm m} $, such that
$ {\rm mod} \, (n, \Delta n_{\rm m}) = 0 $. Then $ \Delta n_{\rm m} $
expresses the number of ``hydrodynamic'' time steps between two
``metric'' time slices where the metric is calculated.

\begin{table}
  \centering
  \caption{Values for the metric refinement parameters
    $ f_{\rm ref} $, $ \rho_{\rm thr} $, $ \Delta t_{\rm ref} $,
    and $ n_{\rm m} $ for regular collapse models and for multiple
    bounce scenarios as used in our simulations (see text for more 
    details).}
  \begin{tabular}{lccccccc}
    \hline
    Collapse &
    $ f_{\rm ref} $ &
    $ \rho_{\rm thr} $ &
    $ \Delta t_{\rm ref} $ &
    $ n_{\rm m}^{(1)} $ &
    $ n_{\rm m}^{(2)} $ &
    $ n_{\rm m}^{(3)} \rule{0 em}{1.1 em} $ \\
    & &
    $ [10^{14} {\rm\ g\ cm}^{- 3}] $ &
    [ms] \\
    \hline
    Regular  & \wz0.1 & \wz0.2 & 5.0 & 100 & 10 &  50 \\
    Multiple &   0.02 &   0.04 & --- & 100 & 10 & --- \\
    \hline
  \end{tabular}
  \label{tab:metric_refinement_values}
\end{table}

During a simulation, the metric resolution parameter should vary with
time in a way which adapts to the specific phase of the collapse
scenario. We split the core collapse evolution into three different
phases: The pre-bounce phase lasts from the start of the evolution
until the maximum density on the grid $ \rho_{\rm max} $ reaches a
threshold value, which is a specified fraction of the nuclear density:
$ \rho_{\rm thr} \equiv f_{\rm ref} \rho_{\rm nuc} $. During that
phase the metric is calculated every $ n_{\rm m}^{(1)} $
``hydrodynamic'' time step. The subsequent bounce phase lasts for a
time $ \Delta t_{\rm ref} $, and the metric is solved on every
$ n_{\rm m}^{(2)} $th time slice. In the subsequent post-bounce phase,
the metric resolution parameter is set to $ n_{\rm m}^{(3)} $.

In models showing multiple bounces, i.e.\ where the maximum density
has multiple distinct peaks in time often separated by several 10~ms,
the evolution is split into a series of consecutive bounce phases
interspersed with phases where the density is below the threshold
marked by $ \rho_{\rm thr} $. During the bounce phases the metric
resolution is given by $ n_{\rm m}^{(2)} $, and before the first
bounce and in between bounces by $ n_{\rm m}^{(1)} $.

The collapse phases introduced above are sketched in
Fig.~\ref{fig:metric_refinement}, while the actual values for
$ f_{\rm ref} $ and $ \Delta t_{\rm ref} $ used in our simulations are
summarized for the two different collapse types in
Table~\ref{tab:metric_refinement_values}.

\begin{figure}
  \resizebox{\hsize}{!}{\includegraphics{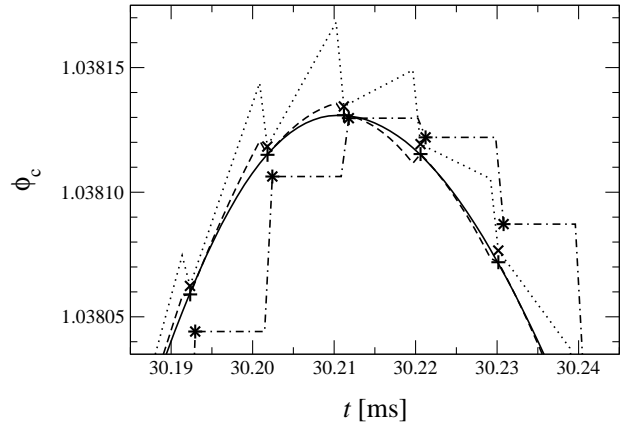}}
  \caption{Approximation of the metric evolution by extrapolation. The
    evolution of the central value of the conformal factor
    $ \phi_{\rm c} $ during core bounce calculated at every time step
    (solid line) is very well approximated by a parabolic
    extrapolation of the metric which is calculated at every 10th time
    step (dashed line, $ + $). If the extrapolation scheme is linear
    (dotted line, $ \times $) or constant (dashed-dotted lines,
    $ * $), the approximation is less accurate. The symbols mark
    the instants when the metric is actually calculated. The model
    used for this plot is A1B3G5 in low grid resolution. For the usual
    resolution of $ 200 \times 30 $ grid points the deviations between
    the different metric extrapolation schemes  are hardly visible.}
  \label{fig:metric_extrapolation}
\end{figure}

In order to approximate the actual evolution of the metric on those
``hydrodynamic'' time slices where it is not calculated, the state of
the metric at old ``metric'' time slices can be used to extrapolate
the metric quantities forward in
time. Fig.~\ref{fig:metric_extrapolation} shows the approximation of
the actual metric calculated at every time slice by a metric which is
calculated at every 10th time step, and is kept constant or is
linearly or parabolically extrapolated in between. Tests with collapse
models demonstrate that our choice of values for the metric resolution
parameters is appropriate to obtain the desired accuracy. Due to the
superior accuracy of the 3rd order parabolic metric extrapolation
scheme we use this method in all our simulations.

%%%%%%%%%%%%%%%%%%%%%%%%%%%%%%%%%%%%%%%%%%%%%%%%%%%%%%%%%%%%%%%%%%%%%%%%%%%%%%%%%%%
%%%%%%%%%%%%%%%%%%%%%%%%%%%%%%%%%%%%%%%%%%%%%%%%%%%%%%%%%%%%%%%%%%%%%%%%%%%%%%%%%%%
%%%%%%%%%%%%%%%%%%%%%%%%%%%%%%%%%%%%%%%%%%%%%%%%%%%%%%%%%%%%%%%%%%%%%%%%%%%%%%%%%%%

\section{Code tests}
\label{sec:tests}

We present now the tests we have used to calibrate our code. It is
worth pointing out that the present code also includes the possibility
of performing simulations of rotational core collapse employing
Newtonian gravity. Many routines and numerical algorithms are common
to both the relativistic and the Newtonian computer code, and by
comparing with previous Newtonian
simulations we have also checked the correct implementation of those
routines. For this purpose we have run some of the models of the
comprehensive sample of \citet{zwerger_97_a} finding excellent
agreement for both the dynamics of the collapse and the
gravitational waveforms. Therefore, whenever we present Newtonian
results below, it is assumed that those have been computed with our
current code. We focus hence on checking our code in the relativistic
regime.

%%%%%%%%%%%%%%%%%%%%%%%%%%%%%%%%%%%%%%%%%%%%%%%%%%%%%%%%%%%%%%%%%%%%%%%%%%%%%%%%%%%
%%%%%%%%%%%%%%%%%%%%%%%%%%%%%%%%%%%%%%%%%%%%%%%%%%%%%%%%%%%%%%%%%%%%%%%%%%%%%%%%%%%

\subsection{Relativistic shock tube tests}
\label{subsec:shock_tubes}

Shock tubes can efficiently test the ability of a code to handle shock
fronts. In a shock tube test, initial data are given for a fluid
system which has two states of constant pressure, density and zero
velocity, initially separated by a boundary. The time-evolution of
these initial data yields a combination of constant states separated
by shocks, contact discontinuities, or rarefaction waves. These tests
allow to compare the numerical results against the analytic solution
\citep{marti_94} in a straightforward way.

\begin{figure}
  \resizebox{\hsize}{!}{\includegraphics{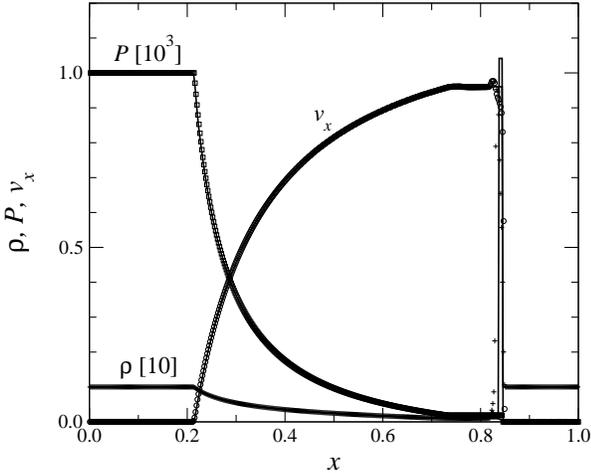}}
  \caption{Numerical (symbols) and analytic (solid line) profiles of
    the (rescaled) density $ \rho $, pressure $ P $, and velocity
    $ v_x $ for the shock tube problem~2.} 
  \label{fig:shock_tube}
\end{figure}

We have simulated the two relativistic shock tube problems analyzed by
\citet{marti_96_a}. The initially constant states are given in
Table~\ref{tab:shock_tubes}. Both cases are computed in a flat
one-dimensional Minkowski spacetime using Cartesian coordinates in the
interval $ x \in [0, 1] $ spanned by 500 equidistant zones. The EoS is
assumed to be that of an ideal fluid with adiabatic index
$ \gamma = 5 / 3 $, and the initial discontinuity is placed at
$ x = 0.5 $. The results for the most stringent problem 2 (a
relativistic blast wave) are displayed in Fig.~\ref{fig:shock_tube},
after an evolution time of $ t_{\rm final} = 0.25 $. The profiles of
the density $ \rho $, the pressure $ P $, and the velocity $ v_x $
accurately reproduce the analytic solution.

\begin{table}
  \centering
  \caption{Initial values of density $ \rho $ and pressure $ P $ for the
    left and right states of the relativistic shock tube problems~1
    and 2 of \citet{marti_96_a}.}
  \begin{tabular}{l|cc|cc|}
    \cline{2-5}
    & \multicolumn{2}{c|}{Problem 1} &
    \multicolumn{2}{c|}{Problem 2} \\
    & left state & right state & left state & right state \\
    \hline
    \multicolumn{1}{|l|}{$ \rho $} &
    10.0 &                     1.0 &      1.0 &          1.0 \\
    \multicolumn{1}{|l|}{$    P $} &
    13.3 & $ 0.66\times 10^{-6} $  & $ 10^3 $ & $ 10^{- 2} $ \\
    \hline
  \end{tabular}
  \label{tab:shock_tubes}
\end{table}

%%%%%%%%%%%%%%%%%%%%%%%%%%%%%%%%%%%%%%%%%%%%%%%%%%%%%%%%%%%%%%%%%%%%%%%%%%%%%%%%%%%
%%%%%%%%%%%%%%%%%%%%%%%%%%%%%%%%%%%%%%%%%%%%%%%%%%%%%%%%%%%%%%%%%%%%%%%%%%%%%%%%%%%

\subsection{Rotating neutron stars}
\label{subsec:rns_tests}

An important test that any axisymmetric hydrodynamics code should pass
is the computation of the evolution of an equilibrium model of a
rotating stellar configuration. If the initial model is computed
accurately, and if the evolution scheme is implemented properly, the
matter and metric profiles of the evolved model should at most
oscillate slightly around their initial equilibrium states, i.e.\ the
model should remain in equilibrium. In particular, for a rotating
model, the rotation profile has to be preserved for as many rotation
periods as possible.

In order to enhance the relativistic effects we have used a rapidly
and uniformly rotating neutron star model with a polytropic matter
distribution (see Table~\ref{tab:rns_model}). This test allows for an
independent check of the metric and the hydrodynamics parts of the
code. If desired, the code can solve only the fluid equations while
keeping the metric fixed (Cowling approximation), or evolve both the
hydrodynamics and the spacetime in a fully coupled evolution. In both
cases we observe that the neutron star remains in equilibrium to high
accuracy for several 10 milliseconds, which corresponds to 10 rotation
periods. In this test run, the metric equations are solved every 200th
hydrodynamic time step, which corresponds to a time interval between
two metric time slices of $0.06 {\rm\ ms}$. This interval is small
compared to both the neutron star's rotation period and the
frequencies of its strongest radial oscillations.

\begin{table}
  \centering 
  \caption{Parameters of the rotating neutron star model used for the
    stability test. $ \rho_{\rm c} $ is the central density,
    $ \gamma $ and $ K $ specify the EoS, $ r_{\rm e} / r_{\rm p} $ is
    the ratio of equatorial to polar radius of the star,
    $ M_{\rm rest/grav} $ are the rest/gravitational masses,
    $ \beta_{\rm rot} $ is the rotation rate,
    $ \Omega / \Omega_{\rm K} $ is the ratio of the star's angular
    velocity to the Keplerian angular velocity and $ T $ is its
    rotation period.}
  \setlength{\tabcolsep}{0.07 cm}
  \begin{tabular}{ccccccccc}
    \hline 
    $ \rho_{\rm c} $ &
    $ \gamma $ & $ K $ &
    $ r_{\rm e} / r_{\rm p} $ &
    $ M_{\rm rest} $ &
    $ M_{\rm grav} $ &
    $ \beta_{\rm rot} $ &
    $ \Omega / \Omega_{\rm K} $ &
    $ T $ \\
    $ [\rho_{\rm nuc}] $ & &
    (cgs) & &
    $ [M_\odot] $ &
    $ [M_\odot] $ &
    [\%] &
    [\%] &
    [ms] \\
    \hline
    3.952 &
    2.0 &
    $ 1.456 \times 10^5 $ &
    0.70 &
    1.756 &
    1.627 &
    7.419 &
    76.0 &
    1.0 \\
    \hline
  \end{tabular}
  \label{tab:rns_model}
\end{table}

\begin{figure}
  \resizebox{\hsize}{!}{\includegraphics{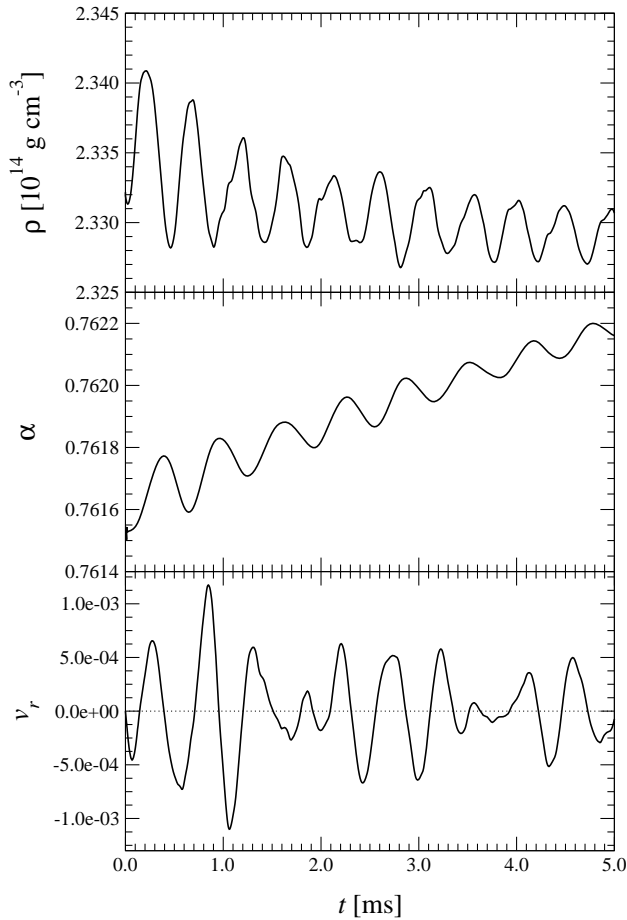}}
  \caption{Time evolution of hydrodynamic and metric quantities of the
    rotating neutron star model of Table~\ref{tab:rns_model}. The
    density $ \rho $ (upper panel), the lapse function $ \alpha $ 
    (middle panel), and the radial velocity $ v_r $ (lower panel) are
    displayed at a radius $ r = 8.77 {\rm\ km} $ in the
    equatorial plane ($ \theta = \pi / 2 $).}
  \label{fig:rns_time_evolution}
\end{figure}

In Fig.~\ref{fig:rns_time_evolution} we show the time evolution of the
density $ \rho $, the lapse $ \alpha $ and the radial velocity
$ v_r = \sqrt{\gamma^{11}} v_1 $ at a radius of 8.77~km in the
equatorial plane for a fully coupled evolution. The small amplitude
oscillations are triggered by numerical discretization errors. The
fact that these oscillations are hardly damped reflects the low
numerical viscosity of the HRSC scheme. By performing a Fourier
transform of these evolved data one can obtain the frequencies of the
different modes of pulsation to high accuracy \citep{font_99_a,
  font_01_a}. Comparisons performed with the results in the Cowling
approximation of \citet{font_01_a} yield a very good agreement for the
mode frequencies.

\begin{figure}
  \resizebox{\hsize}{!}{\includegraphics{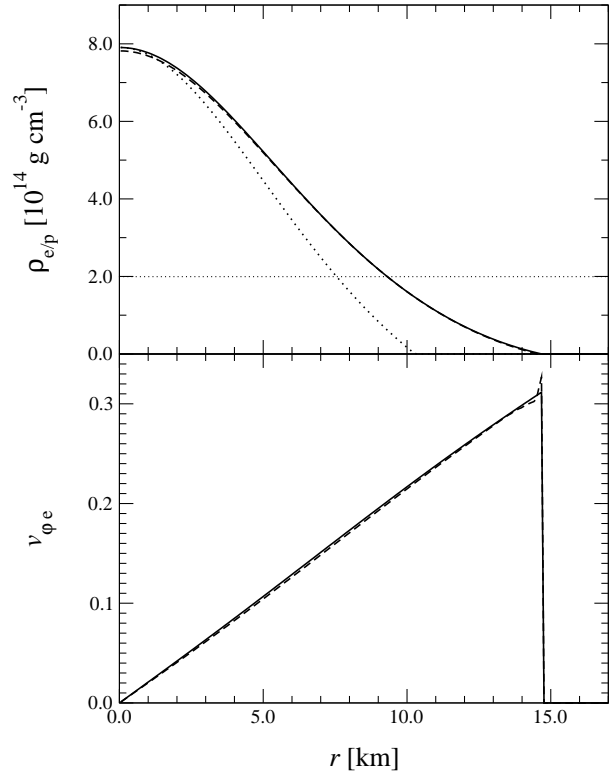}}
  \caption{Density profile (upper panel) and angular velocity profile
    (lower panel) of the rotating neutron star model of
    Table~\ref{tab:rns_model}. The equatorial density profile
    $ \rho_{\rm e} $ at $ t = 5.0 {\rm\ ms} $ (dashed line),
    corresponding to 5 rotation periods is close to the initial
    profile at $ t = 0.0 {\rm\ ms} $ (solid line). The same holds for
    the angular velocity profile $ v_{\varphi\,\rm e} $. The spike in
    the $ t = 5.0 {\rm\ ms} $ profile of $ v_{\varphi\,\rm e} $ close
    to the neutron star's surface is a numerical artifact due to the
    artificial atmosphere. The height of the spike oscillates in time
    with a maximum amplitude as plotted here. Note that the neutron
    star model rotates at more than 30\% of the speed of light at
    the equator. The dotted profile in the upper panel gives the initial
    density along the polar axis $ \rho_{\rm p} $. Due to the rapid
    rotation, the polar radius of the neutron star is only 70\% of its
    equatorial radius. The horizontal dotted line in the density plot
    marks nuclear matter density.}
  \label{fig:rns_profile_evolution}
\end{figure}

As demonstrated in Fig.~\ref{fig:rns_time_evolution}, the code can
keep the equilibrium initial data to a high degree. The small secular
drift observed in $ \rho $ and $ \alpha $ is an artifact of the
numerical scheme and has also been observed by \citet{font_99_a}. This
drift can largely be minimized using a polytropic zero-temperature EoS, which
is a fair assumption for studying small amplitude pulsations of
polytropic stars. The effect of different EoS on the drift is
discussed in \citet{font_01_b}.

In Fig.~\ref{fig:rns_profile_evolution} we plot the radial profiles of
the density $ \rho $ and the rotation velocity
$ v_\varphi = \sqrt{\gamma^{33}} v_3 $ along the equator for the same
neutron star model. Even after 5 rotation periods, the rotation
profile is very close to its initial distribution. Only at the stellar
surface, the angular velocity slightly deviates from its initial shape
due to the interaction with the atmosphere. As shown in
\citet{font_99_a} the use of high-order schemes such as PPM is
essential in preserving the rotational profile.

\begin{figure}
  \resizebox{\hsize}{!}{\includegraphics{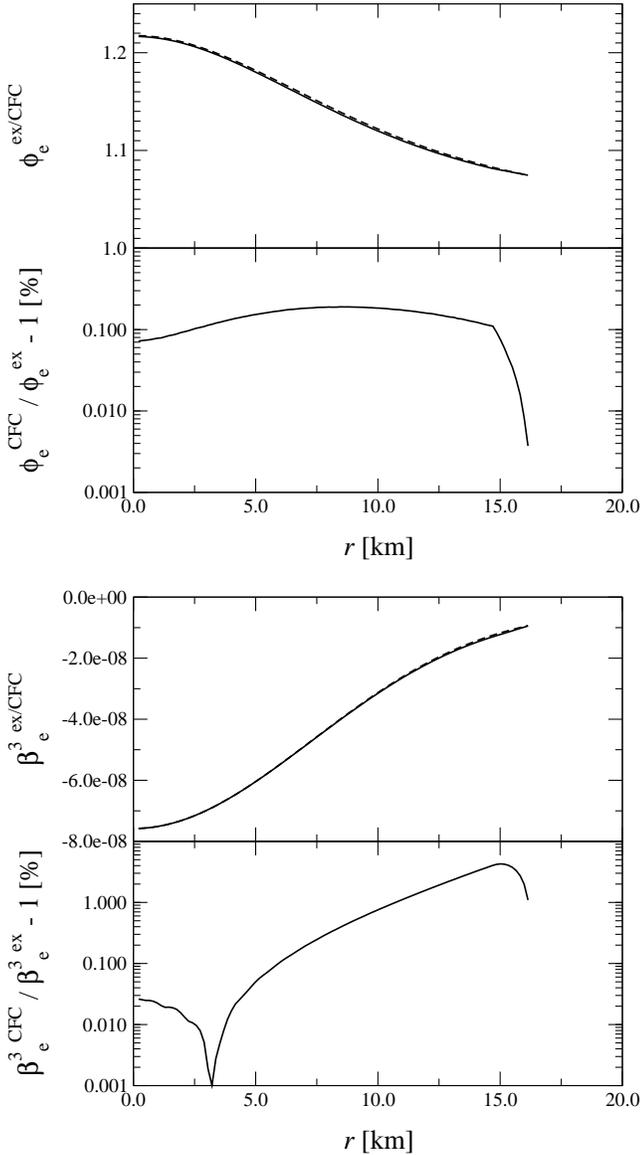}}
  \caption{Quality of the CFC approximation for rapidly rotating
    neutron star initial data. The values for the exact and approximate
    conformal factor $ \phi^{\rm ex}_{\rm e} $ and
    $ \phi^{\rm CFC}_{\rm e} $ (upper top panel), and for the exact
    and approximate shift vector component
    $ \beta^{\rm 3\,ex}_{\rm e} $ and $ \beta^{\rm 3\,CFC}_{\rm e} $
    (lower top panel), both evaluated along the equatorial plane,
    agree very well. In the lower panels the corresponding relative
    deviations are plotted. For the conformal factor the relative
    deviations are less than $ 0.2\% $, while they stay below $ 4\% $
    for the shift vector.}
  \label{fig:rns_cfc_quality_phi_beta}
\end{figure}

Since the neutron star model specified in Table~\ref{tab:rns_model} is
rapidly rotating and thus deviates strongly from spherical symmetry,
its evolution provides a convincing way to test the ability of the CFC
to yield a good approximation of the exact spacetime (see also
Section~\ref{subsec:cfc_quality}). The initial model has been
calculated using the exact metric, Eq.~(\ref{eq:rrs_metric}), without
assuming conformal flatness. The initial data are then matched to the
CFC spacetime. The fact that the evolution code maintains stability
for many rotation periods already proves the validity of the CFC for
rotating neutron star spacetimes. This observation is supported by
Fig.~\ref{fig:rns_cfc_quality_phi_beta} which shows how well the
conformal factor $ \phi^{\rm CFC} $ of the CF metric approximates the
corresponding metric component $ \phi^{\rm ex} $ of the exact
metric. The excellent approximation of the azimuthal component of the
exact shift vector $ \beta^{\rm 3\,ex} $ by the CFC shift vector
component $ \beta^{\rm 3\,CFC} $ is shown in the lower panels. Our
results are in good agreement with those of \citet{cook_96_a}.

%%%%%%%%%%%%%%%%%%%%%%%%%%%%%%%%%%%%%%%%%%%%%%%%%%%%%%%%%%%%%%%%%%%%%%%%%%%%%%%%%%%
%%%%%%%%%%%%%%%%%%%%%%%%%%%%%%%%%%%%%%%%%%%%%%%%%%%%%%%%%%%%%%%%%%%%%%%%%%%%%%%%%%%

\subsection{Spherical core collapse}
\label{subsec:spherical_core_collapse}

The problem of relativistic core collapse for an ideal fluid in
spherical symmetry was first considered by \citet{may_66_a}. Their
code used a Lagrangian formulation, where the coordinates label mass
shells rather than spatial positions and thus comove with the
fluid. Artificial viscosity terms were included to damp out spurious
numerical oscillations behind shock fronts. In this Section we compare
our Eulerian code against a Lagrangian finite difference code
\citep{dimmelmeier_98_a} for a spherically symmetric core collapse.

For this test we have set up a nonrotating equilibrium star as initial
data on a two-dimensional $ (r, \theta) $-grid, with the same EoS and
central density as in the (rotating) initial models listed in
Table~\ref{tab:collapse_models}. During the evolution, we do not use
the hybrid EoS~(\ref{eq:hybrid_eos}), but a simple ideal fluid EoS
given by Eq.~(\ref{eq:ideal_gas_eos}).
    
\begin{figure}
  \resizebox{\hsize}{!}{\includegraphics{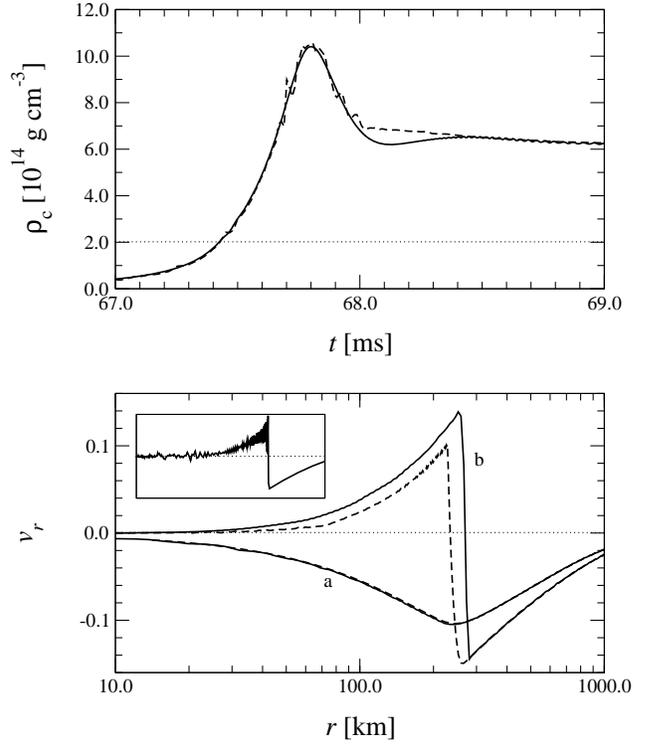}}
  \caption{Spherically symmetric core collapse: Comparison of results
    obtained with our Eulerian HRSC code (solid lines) against results
    from a Lagrangian finite difference code (dashed lines). The time
    evolution of the central density $ \rho_{\rm c} $ (upper panel),
    here plotted at the time of bounce, agrees very well during all
    stages of the collapse. The dotted line in the density plot marks
    nuclear matter density. In the lower panel, the radial velocity
    profiles are plotted at $ t = 60 {\rm\ ms} $ (a) and at
    $ t = 75 {\rm\ ms} $ (b), respectively. During the infall phase
    before core bounce (a), the velocity profiles match very
    well. When the shock front propagates outward (b), the agreement
    is less good due to the more dissipative character of the
    Lagrangian code. The inset shows the growth of unphysical
    oscillations in the velocity profile obtained with the Lagrangian
    code for very low artificial viscosity at $ t = 75 {\rm\ ms} $.}
  \label{fig:spherical_core_collapse}
\end{figure}

The effective stiffening of the EoS at supranuclear matter densities
is modeled by a variable adiabatic index given by the following
relation \citep{van_riper_79_a, romero_96_a}:
\begin{equation}
  \gamma = \left\{
  \begin{array}{ll}
    \displaystyle \gamma_1 \qquad &
    \mbox{for } \rho < \rho_{\rm nuc}, \\
    \displaystyle \gamma_1 +
    \frac{1}{2} \log \frac{\rho}{\rho_{\rm nuc}} \qquad &
    \mbox{for } \rho \ge \rho_{\rm nuc},
  \end{array}
  \right.
  \label{eq:spherical_core_collapse_gamma}
\end{equation}
with $ \gamma_1 = 1.320 $. Besides comparing our results with those of
an independent spherically symmetric code, this test also allows us to
assess the capability of the code to maintain spherical symmetry
throughout the collapse.

The differences in the ability of both codes to handle shocks are
demonstrated in Fig.~\ref{fig:spherical_core_collapse}. In the upper
panel we plot the evolution of the central density around the time of
bounce. We note that due to the relatively soft supranuclear EoS used
for this run the core dives deeply into the potential well reaching
densities as high as $ 5 \rho_{\rm nuc} $. Even during this epoch,
where the highest densities are reached, the evolution computed with
our code follows very accurately the corresponding one obtained with
the May--White code. However, due to its superior effective radial
resolution, the May--White code resolves the drop in the central
density after the maximum peak better, and also produces a smoother
density evolution. Up to the formation of the shock front, both codes
yield equally good results for the radial infall velocity
profiles. Such profiles are plotted in the lower panel of
Fig.~\ref{fig:spherical_core_collapse} at $ t = 60 {\rm\ ms} $ (before
bounce) and $ t = 75 {\rm\ ms} $ (after bounce), respectively. With
the HRSC scheme of our code the shock front is steeper and confined to
fewer grid zones than with the artificial viscosity scheme implemented
in the May--White code. Additionally, the velocity behind the shock is
higher and the shock propagates faster, which is a result of much less
numerical dissipation. We note that if the artificial viscosity is
reduced in the May--White code to cure these negative effects,
unphysical spurious oscillations grow behind the shock front (see
inset in the lower panel of Fig.~\ref{fig:spherical_core_collapse}).

%%%%%%%%%%%%%%%%%%%%%%%%%%%%%%%%%%%%%%%%%%%%%%%%%%%%%%%%%%%%%%%%%%%%%%%%%%%%%%%%%%%
%%%%%%%%%%%%%%%%%%%%%%%%%%%%%%%%%%%%%%%%%%%%%%%%%%%%%%%%%%%%%%%%%%%%%%%%%%%%%%%%%%%

\subsection{Integral quantities -- Conservation of rest-mass and
  angular momentum}
\label{subsec:conservation}

In the 3\,+\,1 formalism of general relativity, the definitions for
the (baryon) rest mass $ M_{\rm rest} $, the proper mass
$ M_{\rm proper} $, the gravitational mass $ M_{\rm grav} $, the
angular momentum $ J_{\rm rot} $, and the rotational mass (energy)
$ M_{\rm rot} $ are \citep{friedman_86_a}:
\begin{equationarray}
  M_{\rm rest} & = & - \int  \rho u_\mu n^\mu {\rm d}V,
  \label{eq:general_m_rest} \\
  M_{\rm proper} & = & - \int  \rho (1 + \epsilon) u_\mu n^\mu {\rm d}V,
  \label{eq:general_m_proper} \\
  M_{\rm grav} & = & - \int  \left( 2 T_{\mu \nu} - g_{\mu \nu} T \right)
  t^\mu n^\nu {\rm d}V,
  \label{eq:general_m_grav} \\
  J_{\rm rot} & = & - \int  T_{\mu \nu} s^\mu n^\nu {\rm d}V,
  \label{eq:general_j_rot} \\
  M_{\rm rot} & = & - \int  T_{\mu \nu} s^\mu n^\nu
  \frac{s_\lambda u^\lambda}{2 n_\kappa u^\kappa} {\rm d}V,
  \label{eq:general_m_rot}
\end{equationarray}%
where $ t^\mu $ and $ s^\mu $ are a time-like and a space-like Killing
vector, respectively. The rotation axis is perpendicular to $ s^\mu $.
For the CF metric in axisymmetry, where $ s^\mu = \hat{e}^3 $ (i.e.\
the unit vector in the $ \varphi $-direction), the above integral
quantities have the following form:
\begin{equationarray}
  M_{\rm rest} & = & - 2 \pi \int r^2 \sin \theta \phi^6
  \rho W {\rm d}r {\rm d}\theta,
  \label{eq:cfc_m_rest} \\
  M_{\rm proper} & = & - 2 \pi \int r^2 \sin \theta \phi^6
  \rho (1 + \epsilon) W {\rm d}r {\rm d}\theta,
  \label{eq:cfc_m_proper} \\
  M_{\rm grav} & = & - 2 \pi \int r^2 \sin \theta \phi^6
  \left( \alpha (2 \rho h W^2 + 2P - \rho h) \right. \nonumber \\
  & & \qquad \quad \left. - 2 \rho h W^2 v_i \beta^i \right)
  {\rm d}r {\rm d}\theta, \label{eq:cfc_m_grav} \\
  J_{\rm rot} & = & - 2 \pi \int r^2 \sin \theta \phi^6
  \rho h W^2 v_3 {\rm d}r {\rm d}\theta,
  \label{eq:cfc_j_rot} \\
  M_{\rm rot} & = & - 2 \pi \int r^2 \sin \theta \phi^6
  \rho h W^2 \frac{v_3 (\alpha v^3 - \beta^3)}{2}
  {\rm d}r {\rm d}\theta.
  \label{eq:cfc_m_rot}
\end{equationarray}%
Note that for $ c = 1 $, the energies have the same units as the
corresponding masses. The potential (mass) energy is given by
$ M_{\rm} = M_{\rm grav} - M_{\rm proper} - M_{\rm rot} $.

This formulation of the above integral quantities is an extension of
the masses and the angular momentum defined in \citet{komatsu_89_a}
for nonzero radial and angular velocity and shift vector components,
applied to a conformally flat metric. It is obvious that the integrand
of the rest mass and the angular momentum can be written as
$ \sqrt{\gamma} D $ and $ \sqrt{\gamma} S_3 $, respectively. These are
the first and fourth component of
Eq.~(\ref{eq:hydro_conservation_equation}) and correspond to the
continuity equation and the angular momentum conservation equation,
respectively. We have used these equations to check the conservation
properties of our numerical code.

First, we have performed tests where we have evolved the hydrodynamic
state vector in a fixed spacetime metric. Additionally, the fluxes at
the outer boundary are explicitely set to zero and no artifical
atmosphere is used. In this situation, we obtain conservation of
the rest mass and the angular momentum of
\begin{equation}
  dM_{\rm rest} \equiv \left| \frac{M_{\rm rest}}{M_{\rm rest\,0}} - 1
  \right| \sim 10^{-15},
\end{equation}
\begin{equation}
  dJ_{\rm rot} \equiv \left| \frac{J_{\rm rot}}{J_{\rm rot\,0}} - 1
  \right| \sim 10^{-15},
  \label{eq:rest_mass_conservation}
\end{equation}
i.e.\ up to machine precision. Here $ M_{\rm rest\,0} $ and
$ J_{\rm rot\,0} $ are the initial total rest mass and total angular
momentum, respectively. In Newtonian gravity, the same level of
conservation is reached.

\begin{figure}
  \resizebox{\hsize}{!}{\includegraphics{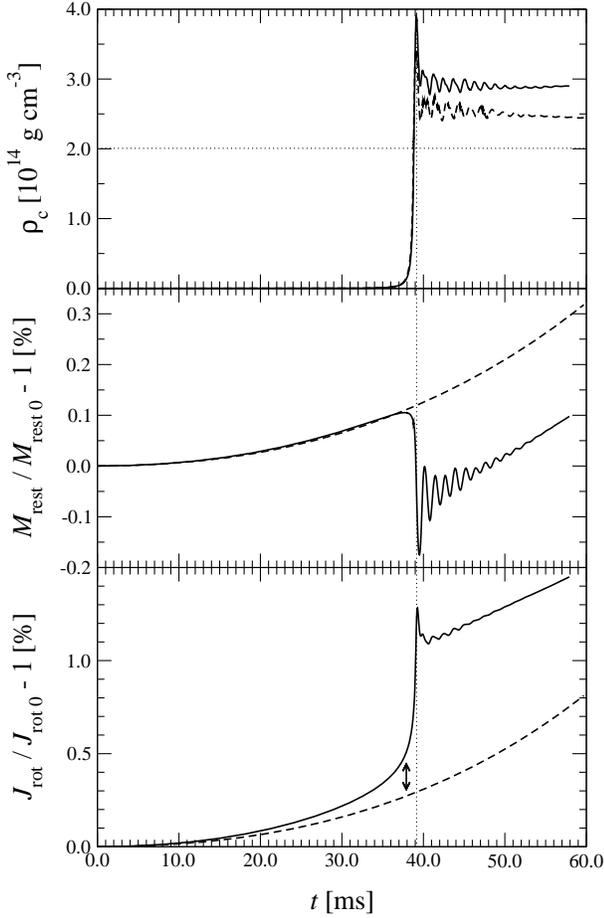}}
  \caption{Conservation of total rest mass and angular momentum in the
    relativistic (solid lines) and Newtonian (dashed lines) simulation
    of model A3B2G4. Until the time of bounce at
    $ t_{\rm b} \approx 39 {\rm\ ms} $ (indicated by the vertical
    dotted line) the relative changes in the total rest mass
    $ M_{\rm rest} $ (middle panel) and angular momentum
    $ J_{\rm rot} $ (lower panel) due to interaction of the stellar
    interior with the artifical atmosphere are small compared to the
    respective initial values $ M_{\rm rest\,0} $ and
    $ J_{\rm rot\,0} $. Nonzero source terms in the evolution equation
    for $ S_3 $ drive additional nonconservation of the angular
    momentum in relativistic gravity (indicated by the arrow). As the
    central density $ \rho_{\rm c} $ (upper panel) and the metric
    volume factor $ \sqrt{\gamma} $ vary most during bounce, a strong
    oscillatory variation of total mass and angular momentum can be
    observed in this phase (associated with the pulsations observed in
    the proto-neutron star). The horizontal dotted line in the density
    plot marks nuclear matter density.}  
  \label{fig:conservation_1}
\end{figure}

\begin{figure}
  \resizebox{\hsize}{!}{\includegraphics{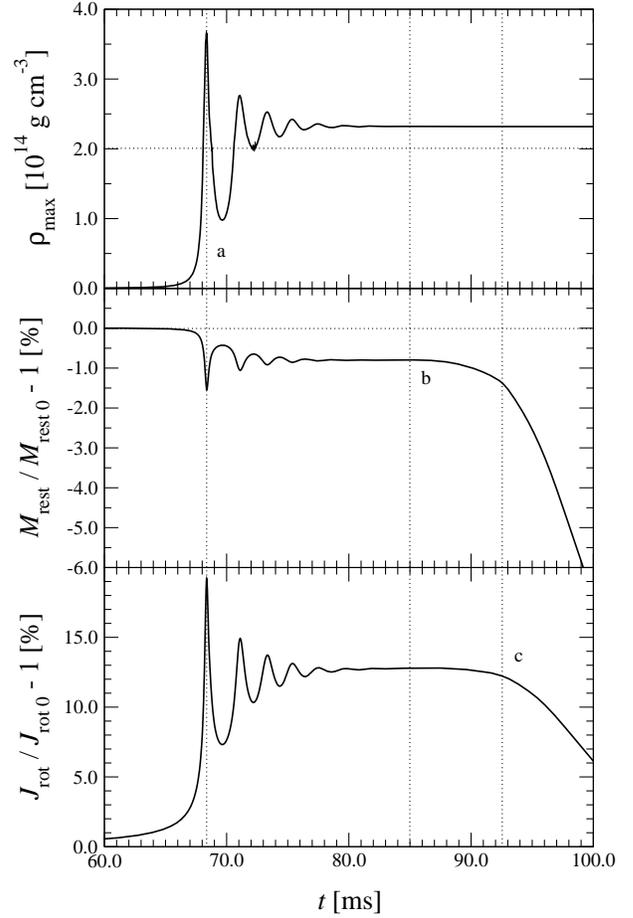}}
  \caption{Influence of the artificial atmosphere on the conservation
    of total rest mass and angular momentum for model A4B2G2. Before
    bounce the change of total mass and angular momentum due to
    interaction of the stellar interior with the atmosphere is
    small. Around the time of bounce at
    $ t_{\rm b} \approx 68.5 {\rm\ ms} $ (a), the maximum density
    $ \rho_{\rm max} $ (upper panel), which is off-center in this
    toroidal model, and the volume element $ \sqrt{\gamma} $ oscillate
    strongly. At $ t \approx 85 {\rm\ ms} $ (b), the shock reaches
    the stellar surface, first at the pole at
    $ R_{\rm b\,p} \approx 1\,200 {\rm\ km} $. This leads to mass loss
    setting in around that time. At $ t \approx 92.5 {\rm\ ms} $ (c),
    the slower shock along the equatorial plane breaks through the
    stellar surface, which is located at
    $ R_{\rm b\,e} \approx 1\,320 {\rm\ km} > R_{\rm b\,p} $. This
    results in a loss of total angular momentum, and also increases
    the mass loss.}
  \label{fig:conservation_2}
\end{figure}

In a dynamic core collapse simulation the above restrictions have to
be abandoned. The nonstationarity of the metric and (in the case of
rotation) the artificial atmosphere prevent a highly accurate
conservation of the integral quantities. In
Fig.~\ref{fig:conservation_1} we plot the behavior of the mass and the
angular momentum in the rotational core collapse model A3B2G4 (see
Table~\ref{tab:collapse_models}). The dashed lines correspond to a
Newtonian simulation and the solid ones are the relativistic
counterparts. In Newtonian gravity the only source for mass and
angular momentum ``accretion'' is the artificial atmosphere. This becomes
visible as a small monotonous increase in $ M_{\rm rest} $ (middle
panel) and $ J_{\rm rot} $ (bottom panel). Until bounce, indicated by
the vertical dotted line, the mass increase in relativistic gravity
exhibits the same behavior. However, due to the nonzero source term in
the relativistic angular momentum equation, $ dJ_{\rm rot} $ is larger
than its Newtonian counterpart already during the infall phase.

Around the time of bounce another effect which prohibits exact
numerical conservation of total rest mass and angular momentum becomes
important. The conservation
equations~(\ref{eq:hydro_conservation_equation}) are formulated for
$ \sqrt{\gamma} \mb{U} $. But in our numerical scheme we evolve only
the state vector $ \mb{U} $ and, for numerical reasons, we treat the
geometric volume term $ \sqrt{\gamma} $ as a source term. Therefore,
for any strong temporal variation of the metric, e.g.\ during
bounce, $ dM_{\rm rest} $ and $ dJ_{\rm rot} $ vary considerably,
too. This is clearly visible in Fig.~\ref{fig:conservation_1}. When
the central density has settled down to an almost constant value, and
the core has reached a new equilibrium configuration, the rates of
increase of $ M_{\rm rest} $ and $ J_{\rm rot} $ are again similar in
relativistic and in Newtonian gravity.

The effects of the artificial atmosphere on the conservation of total
mass and angular momentum can be infered from
Fig.~\ref{fig:conservation_2}. In the highly nonspherical rotational
core collapse model A4B2G2, the shock propagates outward at different
speeds in the polar and the equatorial direction. When the shock reaches the
surface of the star at the pole ($ t\approx 85 {\rm\ ms} $), a
considerable amount of mass, but not much angular momentum is lost to
the atmosphere. The equator is reached by the shock at a later time,
since the shock propagates slower in this direction and the equatorial
radius is larger. Around that time at $ t\approx 92.5 {\rm\ ms} $ the
mass loss curve steepens and significantly more angular momentum is
lost, as the relative contributions of fluid elements to the total
angular momentum is higher near the equatorial plane than at the pole.
Fig.~\ref{fig:conservation_2} also shows the variation of
$ dM_{\rm rest} $ and $ dJ_{\rm rot} $ due to rapid changes of the
volume element $ \sqrt{\gamma} $ around peak maximum density.

%%%%%%%%%%%%%%%%%%%%%%%%%%%%%%%%%%%%%%%%%%%%%%%%%%%%%%%%%%%%%%%%%%%%%%%%%%%%%%%%%%%
%%%%%%%%%%%%%%%%%%%%%%%%%%%%%%%%%%%%%%%%%%%%%%%%%%%%%%%%%%%%%%%%%%%%%%%%%%%%%%%%%%%

\subsection{Quality of the conformally flat condition during rotational
core collapse}
\label{subsec:cfc_quality}

As mentioned in Section~\ref{subsec:rns_tests}, the accuracy of
the CFC approximation has been considered by \citet{cook_96_a}, who
find remarkably good results for rapidly rotating relativistic stars
in equilibrium, typical errors for different variables being smaller
than 5\%. However, the quality of the CFC approximation degrades in
the case of extremely relativistic nonspherical configurations, e.g.\
for rigidly rotating infinitesimally thin disks of dust, which have
been investigated by \citet{kley_99_a}. These authors also point out that
general relativity and the truncated CFC approach are identical up to
the first post-Newtonian order. Although the above comparisons have
been made for stationary configurations, we want to emphasize that the
CFC approximation does not make any explicit assumptions about the
spacetime being static or dynamic. Due to the elliptic character of
the metric
equations~(\ref{eq:metric_equation_1}--\ref{eq:metric_equation_3}),
the spacetime metric is solely determined by the instantaneous
hydrodynamic state. Therefore the deviation of the CF metric from the
exact spacetime metric at a particular time will only be due to the
current deviation of the matter distribution from spherical symmetry.

In this Section we analyze the validity of the CFC in simulations of
rotational core collapse. There are several ways to quantify the
violations of the exact metric equations. Firstly, one could use the CF
metric obtained by solving
Eqs.~(\ref{eq:metric_equation_1}--\ref{eq:metric_equation_3}) to
compute the Einstein tensor $ G_{\mu\nu} $ and to check the Einstein
equations $ G_{\mu\nu} = 8 \pi T_{\mu\nu} $. However, this is a
difficult approach due to the huge number of terms involved.

On the other hand, the ADM metric
equations~(\ref{eq:adm_metric_equation_1}--\ref{eq:adm_metric_equation_4})
are equivalent to the Einstein
equations. As we only use a subset of the ADM equations to calculate
the CF metric, we can use the remaining ones to test the quality of
the CFC approximation. Therefore, we re-insert the CF metric
components into the individual ADM metric equations and compare the
left and right hand sides of the respective equations. We note that
such a procedure has also been used by \citet{gourgoulhon_01_a} in
validating initial data for binary black holes. Comparisons with
Damour's analytic post-Newtonian expansion show satisfactory agreement
in this case up to third post-Newtonian order
\citep{gourgoulhon_01_b}.

If the inserted metric is an exact solution of the Einstein equations,
the numerical inequality between the left and right hand sides of the
ADM equations has to converge to zero with increasing resolution. For
an approximate metric like the CF metric there will remain a nonzero
residual 
\begin{equation}
  r_{q} \equiv
  \left| \frac{{\rm lhs\,} q}{{\rm rhs\,} q} - 1  \right| > 0
  \label{eq:adm_residual}
\end{equation}
in the ADM equations for at least some metric quantities $ q $
irrespective of numerical resolution. Here $ {\rm lhs\,} q $ and
$ {\rm rhs\,} q $ stand for the left and right hand side of the
corresponding ADM metric equation, respectively.

We want to point out that the metric
equations~(\ref{eq:metric_equation_1}--\ref{eq:metric_equation_3}) are
analytically equivalent to the ADM constraint equations. Any solution
of the CF metric equations will also be a solution of the ADM
constraint equations, at least up to the precision of the numerical
scheme.

\begin{figure}
  \resizebox{\hsize}{!}{\includegraphics{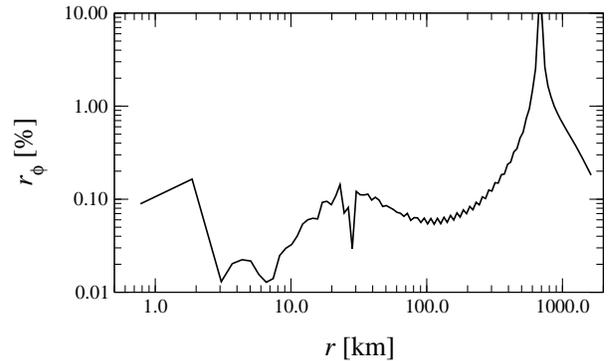}}
  \caption{Numerical equivalence of the left hand side
    (${\rm lhs} \, \phi = \partial_t \phi $) and right hand side
    (${\rm rhs} \, \phi = {1\over6} \phi \nabla_k \beta^k $) of the
    evolution equation for the conformal factor $ \phi $. The residual
    $ r_\phi $ of the evolution equation for $ \phi $, (plotted along
    the equatorial radius at $ t \approx 41 {\rm\ ms} $) is small
    everywhere. Note that the pole at large $ r $ is due to a
    vanishing denominator in Eq.~(\ref{eq:dt_phi_residual}).}
  \label{fig:phi_t_relation}
\end{figure}

The evolution equations for the three-metric $ \gamma_{ij} $,
Eq.~(\ref{eq:adm_metric_equation_1}), are also identically
fulfilled for the off-diagonal elements of $ \gamma_{ij} $, because
for a CF metric they transform into the definition of the extrinsic
curvature, Eq.~(\ref{eq:definition_of_extrinsic_curvature}). As the
three-metric $ \gamma_{ij} $ is conformally flat, of the six evolution
equations~(\ref{eq:adm_metric_equation_1}) there remains only the
evolution equation for the conformal factor $ \phi $,
Eq.~(\ref{eq:phi_t_relation}) which must be monitored. We have found
that the residual of this equation,
\begin{equation}
  r_{\phi} =
  \left| \frac{{\rm lhs\,} \phi}{{\rm rhs\,} \phi} - 1  \right| =
  \left| \frac{\partial_t \phi}{{1\over6} \phi \nabla_k \beta^k} - 1 \right|,
  \label{eq:dt_phi_residual}
\end{equation}
is very close to zero in various test situations including rotational
core collapse. For model A3B4G3 the radial profile of $ r_{\phi} $ is
plotted in Fig.~\ref{fig:phi_t_relation} shortly before maximum
density is reached. Except where a pole develops due to a zero
denominator in Eq.~(\ref{eq:dt_phi_residual}), $ r_{\phi} $ is
everywhere less than 0.2\%, particularly in the dense interior.
Hence, we infer that the ADM evolution equations for the diagonal
elements of $ \gamma_{ij} $ are solved to high degree of accuracy by
the CF metric as well.

The ADM evolution equations for the extrinsic curvature provide
another way of checking the accuracy of the CFC approximation. As a
consequence of the maximal slicing condition, $ K=0 $, the sum of the
left and right hand sides of the equation for the diagonal elements of
the extrinsic curvature are identical up to numerical discretization
errors at all grid points in any evolution. Furthermore, in spherical
symmetry, the left and right hand sides of all six evolution equations
for $ K_{ij} $ agree, because the CF metric is an exact solution of
the spacetime in this case.

\begin{figure}
  \resizebox{\hsize}{!}{\includegraphics{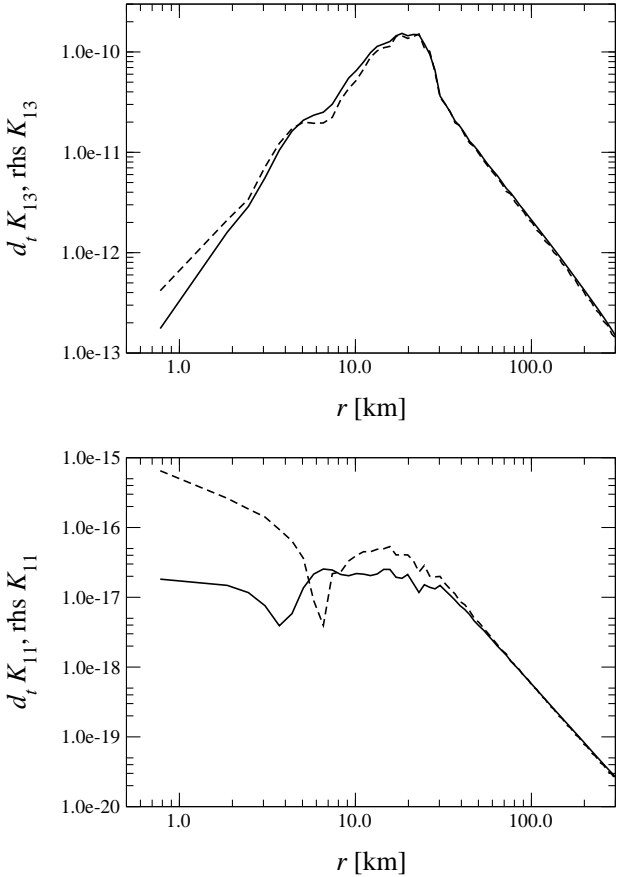}}
  \caption{Effect of the CFC approximation on the ADM evolution
    equation for $ K_{13} $ (upper panel) and $ K_{11} $ (lower panel)
    in the rotational core collapse model A3B4G3. In the case of
    $ K_{13} $, the equatorial radial profiles of the left hand side
    (solid lines) and of the right hand side (dashed lines) of the ADM
    evolution equation agree very well down to small radii where
    relativistic effects are most important. For $ K_{11} $ the terms
    $ - \nabla_i \nabla_j \alpha $ and $ \alpha R_{ij} $ are nonzero.
    This causes the non-negligible misalignment between both sides of
    that equation (bottom panel). The comparison is done at
    $ t = 41 {\rm\ ms} $, which is about 1~ms before bounce.}
  \label{fig:extrinsic_curvature_core_collapse}
\end{figure}

A violation of the evolution equations for the diagonal elements of
$ K_{ij} $, Eq.~(\ref{eq:adm_metric_equation_2}), occurs in all highly
nonspherical and relativistic spacetimes, e.g.\ in the core collapse
model A3B4G3 (Fig.~\ref{fig:extrinsic_curvature_core_collapse}). The
agreement between both sides of the equations is good at low densities
at the beginning of the infall phase. However, when high densities are
encountered in the center during and after bounce the misalignment
becomes as prominent as for the rotating equilibrium neutron star
model of Section~\ref{subsec:rns_tests} (bottom panel of
Fig.~\ref{fig:extrinsic_curvature_core_collapse}). On the other hand,
the evolution equations for off-diagonal components are solved to high
accuracy. This is demonstrated by he very good agreement of the left
and right hand sides of the equation for $ K_{13}$ even in the center
of the core (upper panel of
Fig.~\ref{fig:extrinsic_curvature_core_collapse}).

These observations raise the question why the left and right hand
sides of some of the evolution equations for the extrinsic curvature
disagree considerably for nonspherical matter distributions. The
reason becomes obvious when analyzing the structure of those
equations. In comprehensive tests computing many dynamic collapse
models we have found that the terms in
Eq.~(\ref{eq:adm_metric_equation_2} containing extrinsic containing
extrinsic curvature components are negligibly small compared to the
other three terms $ - \nabla_i \nabla_j \alpha $, $ \alpha R_{ij} $
and $ - 8 \pi \alpha (S_{ij} - {1\over2} \gamma_{ij} (S - \rho_{\rm H})) $
everywhere on the grid and during all epochs. The relevant terms
depend only on the metric quantities $ \alpha $ (which is fixed by a
gauge condition) and $ \phi $, but not on the shift vector components
$ \beta^i $. This points towards the inability of the CF three-metric
$ \gamma_{ij} $, which has only one nonzero component $ \phi $, to
exactly represent the physical spacetime. On the other hand, we note
that the quality of the shift vector components, which are calculated
directly according to the ADM momentum constraint
equations~(\ref{eq:adm_metric_equation_4}) (which are equivalent to
the metric equation~(\ref{eq:metric_equation_3})), is not measured by
Eq.~(\ref{eq:adm_metric_equation_2}).

Therefore, the misalignment of the evolution equation for the
extrinsic curvature is not a result of strong deviations of the shift
vector components, but is instead due to the restriction imposed onto
the three-metric $ \gamma_{ij} $ by the CFC approximation. This
conjecture is additionally supported by the fact that, as demonstrated
in Fig.~\ref{fig:rns_cfc_quality_phi_beta} in
Section~\ref{subsec:rns_tests}, even for high density, rapidly
rotating neutron stars in equilibrium the value of the shift vector
component $ \beta^3 $ calculated by the CFC
equation~(\ref{eq:metric_equation_3}) remains close to the exact
value.

This also explains the perfect matching of the left and right hand
sides of Eq.~(\ref{eq:adm_metric_equation_2}) for $ K_{13} $ even for
rapidly rotating and collapsing models. For this component the
problematic terms $ - \nabla_i \nabla_j \alpha $ and $ \alpha R_{ij} $
vanish. However, for $ K_{12} $ and $ K_{23} $ the deviations are more
pronounced, since the polar velocity $ v_2 $ is very close to zero in
our collapse models. As a result the left and right hand sides of
these evolution equations are dominated by truncation errors.

%%%%%%%%%%%%%%%%%%%%%%%%%%%%%%%%%%%%%%%%%%%%%%%%%%%%%%%%%%%%%%%%%%%%%%%%%%%%%%%%%%%
%%%%%%%%%%%%%%%%%%%%%%%%%%%%%%%%%%%%%%%%%%%%%%%%%%%%%%%%%%%%%%%%%%%%%%%%%%%%%%%%%%%
%%%%%%%%%%%%%%%%%%%%%%%%%%%%%%%%%%%%%%%%%%%%%%%%%%%%%%%%%%%%%%%%%%%%%%%%%%%%%%%%%%%

\section{Conclusions}
\label{sec:conclusions}

We have presented a detailed description of a new axisymmetric general
relativistic code for rotational core collapse which solves the
coupled system of Einstein equations and fluid equations. Those
equations are integrated adopting the ADM 3\,+\,1 formalism to foliate
the spacetime into a set of non-intersecting spacelike hypersurfaces. In
order to simplify the metric equations we approximate the general
metric using the conformal flatness condition (CFC) introduced by
\citet{wilson_96_a}, whereby the spatial components of the metric are
set equal to the flat three-metric times a conformal factor which depends
on the coordinates. The CFC implies a considerable simplification of
the ADM 3\,+\,1 metric equations which reduce to a set of five coupled
non-linear elliptic equations for the metric components. We have
carefully analyzed the applicability and quality of the CFC, showing
that this approximation is appropriate for simulations of rotational
core collapse.
 
We have also demonstrated that our general relativistic hydrodynamics
solver accurately passes a set of difficult test problems including
special relativistic shock tubes, the preservation of the rotation
profile and of the equilibrium of rapidly and differentially rotating
relativistic polytropes, and the spherical collapse of a relativistic
polytrope. The solver, which is based on a high-resolution shock
capturing scheme with an approximate Riemann solver, also conserves
accurately rest-mass and angular momentum in dynamic spacetimes.

The code described in this work was developed with the aim to simulate
the general relativistic axisymmetric collapse of rotating stellar
cores. To this end we have computed and presented here a large set of
relativistic rotating precollapse equilibrium models with different
amounts and distributions of angular momentum. Using these initial
models we have performed a series of collapse simulations and
calculated the resulting gravitational wave signal. The results of
this first relativistic study of rotational core collapse and
of its gravitational wave signature are discussed in an accompanying
paper \citep{dimmelmeier_02_b}.

%%%%%%%%%%%%%%%%%%%%%%%%%%%%%%%%%%%%%%%%%%%%%%%%%%%%%%%%%%%%%%%%%%%%%%%%%%%%%%%%%%%
%%%%%%%%%%%%%%%%%%%%%%%%%%%%%%%%%%%%%%%%%%%%%%%%%%%%%%%%%%%%%%%%%%%%%%%%%%%%%%%%%%%
%%%%%%%%%%%%%%%%%%%%%%%%%%%%%%%%%%%%%%%%%%%%%%%%%%%%%%%%%%%%%%%%%%%%%%%%%%%%%%%%%%%

\begin{acknowledgements}
  We are grateful to Y.~Eriguchi for providing us with a code for
  constructing the intial models. It is a pleasure to acknowledge
  J.~M$ ^{\underline{\mbox{\scriptsize a}}} $~Ib\'a\~nez and
  G.~Sch\"afer for very useful comments and suggestions. We also want
  to thank M.~Rampp for insightful discussions about the sweeping
  method for sparse linear problems. J.A.F.\ acknowledges support from
  a Marie Curie fellowship by the European Union
  (MCFI-2001-00032) and from the Spanish Ministerio de Ciencia y
  Tecnolog\'\i a (AYA2001-3490-C02-01). All computations were
  performed on the NEC SX-5/3C supercomputer at the Rechenzentrum
  Garching.
\end{acknowledgements}

\bibliographystyle{aa}
\bibliography{aa_paper}

\appendix

%%%%%%%%%%%%%%%%%%%%%%%%%%%%%%%%%%%%%%%%%%%%%%%%%%%%%%%%%%%%%%%%%%%%%%%%%%%%%%%%%%%
%%%%%%%%%%%%%%%%%%%%%%%%%%%%%%%%%%%%%%%%%%%%%%%%%%%%%%%%%%%%%%%%%%%%%%%%%%%%%%%%%%%
%%%%%%%%%%%%%%%%%%%%%%%%%%%%%%%%%%%%%%%%%%%%%%%%%%%%%%%%%%%%%%%%%%%%%%%%%%%%%%%%%%%

\section{Recovery of the primitive variables}
\label{app:recovery}

For an ideal gas EoS a suitable method to recover the primitive
quantities from the state vector was introduced by
\citet{marti_91_a}. It has been described in detail in
\citet{marti_96_a}. We have extended this method to be adequate for
our hybrid EoS~(\ref{eq:hybrid_eos}).

We calculate a function
$ f (P^*) \equiv P (\rho^*, \epsilon^*) - P^* $, where $ \rho^* $ and
$ \epsilon^* $ depend on the conserved quantities and $ P^* $ only (in
all relations involving the pressure we replace $ P $ by $ P^* $):
\begin{equationarray}
  f (P^*) & = & K \left( D \sqrt{X} \right)^{\gamma}
  \left( 1 - \frac{\gamma_{\rm th} - 1}{\gamma - 1} \right) 
  \nonumber \\
  & & + (\gamma_{\rm th} - 1) X
  \left[ \tau + D \left( 1 - \frac{1}{\sqrt{X}} \right) +
  P^* \left( 1 - \frac{1}{X} \right) \right] \nonumber \\
  & & - \frac{(\gamma_{\rm th} - 1)(\gamma - \gamma_1)}
  {(\gamma_1 - 1)(\gamma_2 - 1)} K \rho_{\rm nuc}^{\gamma_1 - 1}
  D \sqrt{X} - P^*,
  \label{eq:recovery_f}
\end{equationarray}%
where we use the auxiliary quantity
$ X \equiv 1 - S^2 / (\tau + D + P^*)^2 $, and $ S = S_i S^i $.
The new pressure is iteratively computed using the Newton--Raphson
method:
\begin{equation}
  P^{*\rm\,new} = P^* - f (P^*)
  \left( \frac{{\rm d}f (P^*)}{{\rm d}P^*} \right)^{-1},
  \label{eq:recovery_new_p}
\end{equation}
where the derivative of $ f (P^*) $ is given by
\begin{equationarray}
  \frac{{\rm d}f (P^*)}{{\rm d}P^*} & = &
  \frac{K \left( D \sqrt{X} \right)^{\gamma} S^2
  \left(1 - \frac{\gamma_{\rm th} - 1}{\gamma - 1} \right)}
  {X (\tau + D + P^*)^3} 
  \nonumber \\
  & & + \frac{2 (\gamma_{\rm th} - 1) S^2
  \left[ \tau \!+\! D \left( 1 \!-\! \frac{1}{\sqrt{X}} \right) \!+\!
  P^* \left( 1 \!-\! \frac{1}{X} \right) \right]}{(\tau + D + P^*)^3}  
  \nonumber \\
  & & + (\gamma_{\rm th} - 1) X \left[ \frac{D S^2}{X^{3 / 2}
  (\tau + D + P^*)^3} + 1 \right.
  \nonumber \\
  & & \left. \qquad \qquad \qquad \, - \frac{1}{X} +
  \frac{2 P^* S^2}{X^2 (\tau + D + P^*)^3} \right] 
  \nonumber \\
  & & - \frac{(\gamma_{\rm th} - 1)(\gamma \!-\! \gamma_1) K
  \rho_{\rm nuc}^{\gamma_1 \!-\! 1} D S}
  {(\gamma_1 \!-\! 1)(\gamma_2 \!-\! 1)
  \sqrt{X} (\tau + D + P^*)^3} - 1.
  \label{eq:recovery_df_dp}
\end{equationarray}%
During each iteration, the primitive quantities and the Lorentz
factor, which are needed in the EoS, are updated using the following
relations:
\begin{equationarray}
  v_i^* & = & \frac{S_i}{\tau + D + P^*},
  \label{eq:recovery_primitive_update_v} \\
  W^* & = & \frac{1}{\sqrt{1 - v_i^* v^{i*}}},
  \label{eq:recovery_primitive_update_w} \\
  \rho^* & = & \frac{D}{W^*},
  \label{eq:recovery_primitive_update_rho} \\
  \epsilon^* & = & \frac{\tau + D (1 - W^*) + P^* (1 - W^{*2})}{D W^*}.
  \label{eq:recovery_primitive_update_eps}
\end{equationarray}%
The iteration is continued until convergence is reached.

%%%%%%%%%%%%%%%%%%%%%%%%%%%%%%%%%%%%%%%%%%%%%%%%%%%%%%%%%%%%%%%%%%%%%%%%%%%%%%%%%%%
%%%%%%%%%%%%%%%%%%%%%%%%%%%%%%%%%%%%%%%%%%%%%%%%%%%%%%%%%%%%%%%%%%%%%%%%%%%%%%%%%%%
%%%%%%%%%%%%%%%%%%%%%%%%%%%%%%%%%%%%%%%%%%%%%%%%%%%%%%%%%%%%%%%%%%%%%%%%%%%%%%%%%%%

\section{Explicit form of the metric equations}
\label{app:metric_equations}

Using the notation of Eqs.~(\ref{eq:metric_solution_vector})
and~(\ref{eq:metric_equations_vector}),
the metric equations read (with $ u_{,f} = \partial_f u $,
$ u_{,fg} = \partial_f \partial_g u $):
\begin{equationarray}
  f^1 (u^k) & = & u^1_{,rr} + \frac{2 u^1_{,r}}{r} +
  \frac{u^1_{,\theta\theta}}{r^2} + \frac{\cot \theta}{r^2} u^1_{,\theta} 
  \nonumber \\
  & & + 2 \pi (u^1)^5 (\rho h W^2 - P)
  \nonumber \\
  & & + \frac{(u^1)^7}{48 (u^2)^2} \Biggl[ 3 r^2 \sin^2 \theta (u^5_{,r})^2 +
  \frac{3 (u^3_{,\theta})^2}{r^2} + \frac{4 (u^3)^2}{r^2} \Biggr.
  \nonumber \\
  & & + 4 (u^3_{,r})^2 + 3 r^2 (u^4_{,r})^2 +
  3 \sin^2 \theta (u^5_{,\theta})^2 - \frac{8 u^3 u^3_{,r}}{r}
  \nonumber \\
  & & + 6 u^3_{,\theta} u^4_{,r} + \frac{4 u^3 u^4_{,\theta}}{r} -
  4 u^3_{,r} u^4_{,\theta} + 4 (u^4_{,\theta})^2 
  \nonumber \\
  & & + \frac{4 \cot \theta u^3 u^4}{r} - 4 \cot \theta u^4 u^3_{,r} 
  \nonumber \\
  & & - \Biggl. 
  4 \cot \theta u^4 u^4_{,\theta} + 4 \cot^2 \theta (u^4)^2 \Biggr] = 0,
  \label{eq:metric_equations_u_vector_1} \\
  \nonumber \\
  f^2 (u^k) & = & u^2_{,rr} + \frac{2 u^2_{2,r}}{r} +
  \frac{u^2_{,\theta\theta}}{r^2} + \frac{\cot \theta}{r^2} u^2_{,\theta} 
  \nonumber \\
  & & - 2 \pi (u^1)^4 u^2 (\rho h (3 W^2 - 2) + 5 P) 
  \nonumber \\
  & & - \frac{7 (u^1)^7}{48 (u^2)^2} \Biggl[ 3 r^2 \sin^2 \theta (u^5_{,r})^2 +
  \frac{3 (u^3_{,\theta})^2}{r^2} + \frac{4 (u^3)^2}{r^2} \Biggr.
  \nonumber \\
  & & + 4 (u^3_{,r})^2 + 3 r^2 (u^4_{,r})^2 +
  3 \sin^2 \theta (u^5_{,\theta})^2 - \frac{8 u^3 u^3_{,r}}{r}  
  \nonumber \\
  & & + 6 u^3_{,\theta} u^4_{,r} + \frac{4 u^3 u^4_{,\theta}}{r} -
  4 u^3_{,r} u^4_{,\theta} + 4 (u^4_{,\theta})^2
  \nonumber \\
  & & + \frac{4 \cot \theta u^3 u^4}{r} - \Biggl. 4 \cot \theta u^4 u^3_{,r} 
  \nonumber \\
  & & - 4 \cot \theta u^4 u^4_{,\theta} + 4 \cot^2 \theta (u^4)^2 \Biggr] = 0,
  \label{eq:metric_equations_u_vector_2} \\
  \nonumber \\
  f^3 (u^k) & = & u^3_{,rr} + \frac{2 u^3_{,r}}{r} +
  \frac{u^3_{,\theta\theta}}{r^2} + \frac{\cot \theta}{r^2} u^3_{,\theta} 
  \nonumber \\
  & & - \frac{2 u^3}{r^2} - \frac{2 \cot \theta u^4}{r} -
  \frac{2 u^4_{,\theta}}{r} - \frac{16 \pi u^2 S_1}{u^1}
  \nonumber \\
  & & - \frac{1}{u^2} \Biggl[ \left( u^2_{,\theta} -
  \frac{7 u^2 u^1_{,\theta}}{u^1} \right) \left( u^4_{,r} +
  \frac{u^3_{,\theta}}{r^2} \right) \Biggr.
  \nonumber \\
  & & + \frac{2}{3} \left( u^2_{,r} - \frac{7 u^2 u^1_{,r}}{u^1} \right) 
  \biggl( 2 u^3_{,r} - u^4_{,\theta} - \frac{2 u^3}{r} \biggr.
  \nonumber \\
  & & \Biggl. \biggl. - \cot \theta u^4 \biggr) \Biggr] + \frac{1}{3}
  \Biggl( u^3_{,rr} + u^4_{,r\theta} + \frac{2 u^3_{,r}}{r} \Biggr.
  \nonumber \\
  && \Biggl. - \frac{2 u^3}{r^2} + \cot \theta u^4_{,r} \Biggr) = 0,
  \label{eq:metric_equations_u_vector_3} \\
  \nonumber \\
  f^4 (u^k) & = & u^4_{,rr} + \frac{4 u^4_{,r}}{r} +
  \frac{u^4_{,\theta\theta}}{r^2} +
  \frac{\cot \theta}{r^2} u^4_{,\theta} + \frac{2 u^3_{,\theta}}{r^3}
  \nonumber \\
  & & + \frac{(1 - \cot \theta) u^4}{r^2} - \frac{16 \pi u^2 S_2}{r^2 u^1} 
  \nonumber \\
  & & - \frac{1}{u^2} \Biggl[ \left( u^2_{,r} -
  \frac{7 u^2 u^1_{,r}}{u^1} \right) \left( u^4_{,r} +
  \frac{u^1_{,\theta}}{r^2} \right) \Biggr.
  \nonumber \\
  & & + \frac{2}{3 r^2} \left( u^2_{,\theta} -
  \frac{7 u^2 u^1_{,\theta}}{u^1} \right)
  \biggl( - u^3_{,r} + 2 u^4_{,\theta} \biggr.
  \nonumber \\ 
  & & \Biggl. \biggl. + \frac{u^3}{r} - \cot \theta u^4 \biggr) \Biggr] +
  \frac{1}{3 r^2} \Biggl( u^3_{,r\theta} + u^4_{,\theta\theta} \Biggr.
  \nonumber \\
  & & \Biggl. + \frac{2 u^3_{,\theta}}{r} -
  \frac{u^4}{\sin^2 \theta} \Biggr) = 0,
  \label{eq:metric_equations_u_vector_4} \\
  \nonumber \\
  f^5 (u^k) & = & u^5_{,rr} + \frac{4 u^5_{,r}}{r} +
  \frac{u^5_{,\theta\theta}}{r^2} + \frac{3 \cot \theta}{r^2} u^5_{,\theta} 
  \nonumber \\
  & & - \frac{16 \pi u^2 S_3}{r^2 \sin^2 \theta u^1} -
  \frac{1}{u^2} \left[ \left( u^2_{,r} -
  \frac{7 u^2 u^1_{,r}}{u^1} \right) u^5_{,r} \right.
  \nonumber \\
  & & + \left. \frac{1}{r^2} \left( u^2_{,\theta} -
  \frac{7 u^2 u^1_{,\theta}}{u^1} \right) u^5_{,\theta} \right] = 0.
  \label{eq:metric_equations_u_vector_5}
\end{equationarray}%

\end{document}